\documentclass[11pt,a4paper]{article}
\usepackage{lmodern}
\usepackage[T1]{fontenc}
\usepackage[latin9]{inputenc}
\usepackage{color}
\usepackage{bbding}
\usepackage{multirow}
\usepackage{amsmath}
\usepackage{amsthm}
\usepackage{mathtools}
\usepackage{graphicx}
\usepackage{rotating}
\usepackage{microtype}
\usepackage{float}
\usepackage{xcolor}
\usepackage[unicode=true,pdfusetitle,
 bookmarks=true,bookmarksnumbered=false,bookmarksopen=false,
 breaklinks=false,pdfborder={0 0 1},backref=false,colorlinks=true]
 {hyperref}
\hypersetup{
 allcolors=blue}

\makeatletter

\pdfpageheight\paperheight
\pdfpagewidth\paperwidth

\providecommand{\tabularnewline}{\\}

\usepackage{amssymb}

\makeatother

\usepackage[style=authoryear, backend=biber]{biblatex}
\DeclareMathOperator*{\argmax}{arg\,max}

\begin{document}
\title{Joint leave-group-out cross-validation in Bayesian spatial models}
\author{Alex Cooper\\
    \texttt{alexander.cooper@monash.edu}
    \and
    Aki Vehtari\\
    \texttt{Aki.Vehtari@aalto.fi}
    \and
    Catherine Forbes\\
    \texttt{catherine.forbes@monash.edu}}
\renewbibmacro{in:}{} % remove "In:" in bibliography

\maketitle

\newcommand{\mss}{\ensuremath{\widehat{\mathrm{elpd}}_{CV}\left(M_A, M_B\,|\,y\right)}}
\newcommand{\p}{\mathrm{p}\!}%
\newcommand{\ytest}{\tilde{y}}%
\newcommand{\ytrain}{\mathring{y}}%
\newcommand{\Ntrain}{\mathring{N}}%
\newcommand{\Ntest}{\tilde{N}}%
\newcommand{\Rtrain}{\mathring{R}_{\rho}}%
\newcommand{\Rtest}{\tilde{R}_{\rho}}%
\newcommand{\N}{\mathcal{N}\!}%

\abstract{
Cross-validation (CV) is a widely-used method of predictive assessment based on repeated model fits to different subsets of the available data. CV is applicable in a wide range of statistical settings. However, in cases where data are not exchangeable, the design of CV schemes should account for suspected correlation structures within the data. CV scheme designs include the selection of left-out blocks and the choice of scoring function for evaluating predictive performance.

This paper focuses on the impact of two scoring strategies for block-wise CV applied to spatial models with Gaussian covariance structures. We investigate, through several experiments, whether evaluating the predictive performance of blocks of left-out observations jointly, rather than aggregating individual (pointwise) predictions, improves model selection performance.  Extending recent findings for data with serial correlation (such as time-series data), our experiments suggest that joint scoring reduces the variability of CV estimates, leading to more reliable model selection, particularly when spatial dependence is strong and model differences are subtle.}

\section{Introduction}

Tobler's (\citeyear{tobler1970computer}) first law of geography asserts
that data generated by economic and ecological phenomena are usually
spatial in nature: physically close observations are more
similar to distant ones. Models of these data must therefore deal
with spatial correlation structures \parencite{anselin1988spatial},
and in particular, model selection must account for spatial structure
lest it overestimate predictive ability \parencite{telfordEvaluationTransferFunctions2009,robertsCrossvalidationStrategiesData2017}.

In this paper, we consider model selection procedures using cross-validation
\parencite[CV;][]{vehtariBayesianModelAssessment2002,arlotSurveyCrossvalidationProcedures2010}
for spatial models with Gaussian Markov random field \parencite[GMRF;][]{rueGaussianMarkovRandom2005}
covariance structures. CV is a popular model selection technique that
assesses predictive performance using repeated model re-fits
to data subsets (\emph{folds}). 
CV requires that test and validation sets be reasonably independent \parencite{hastie2009elements,arlotSurveyCrossvalidationProcedures2010}.
This is challenging to guarantee, especially when unknown
dependence relations exist between observations, as is common in
spatial data analysis.

To deal with spatial structure in CV, the analyst must choose data
splits (the `blocking design') and a strategy for
numerically evaluating (scoring) predictions so that the resulting
estimates of predictive ability are indicative of the predictive task
at hand. A large literature considers blocking designs for spatial
problems: see \textcite{robertsCrossvalidationStrategiesData2017}
and \textcite{mahoneyAssessingPerformanceSpatial2023} for accessible
summaries; a number of studies have found that CV performs well when each fold leaves
out contiguous groups of observations \parencite[see also][]{adinAutomaticCrossvalidationStructured2023}.
However, the appropriate scoring strategy for spatial problems has
received less attention.

In this paper, we study the computation of predictive scores 
for blocks of left-out observations. Where models are capable of producing
multivariate probabilistic predictions (i.e. where the predictions
are distribution-valued), scores can either be computed \emph{pointwise}
as a set of univariate predictions and aggregated, or otherwise \emph{jointly
}where the left-out set forms a single multivariate prediction. Recent
work on CV for Bayesian models of time-series data has found that
joint scoring methods outperform pointwise predictions by reducing
the relative variability of the resulting CV estimates \parencite{cooperCrossvalidatoryModelSelection2024}
when strong serial dependence is present. The natural question then
arises: does this phenomenon also appear under spatial dependence?

CV is a statistical procedure subject to sampling variability. Unfortunately,
 it is challenging to characterize the behavior of CV estimators
(for example, there is usually no unbiased estimator for the variance of CV estimators
\cite{bengioNoUnbiasedEstimator2004a,sivulaUnbiasedEstimatorVariance2020}). In this paper, we extend
the analysis of \textcite{cooperCrossvalidatoryModelSelection2024}
and use a simulation-based approach to analyze the frequency properties
of CV model selection for several spatial models with Gaussian covariance
structure. This class includes several workhorse models popular in
econometric analysis, spatial autoregressions \parencite[SAR;][]{anselin1988spatial,hootenSimultaneousAutoregressiveSAR2019,verhoefSpatialAutoregressiveModels2018} (also known as spatial lag models, SLM)
and conditional autoregressive \parencite[CAR;][]{cressie2015statistics} models.

Our results apply to model selection applications where CV finds only small
differences between models.
Of course, in many settings, the scoring approach does not matter because
the better model shines through regardless of the scoring approach.
In others, no CV design will ever be able to clearly identify a better model
with the available data. We are interested in the third, marginal case between
these two extremes, where a moderate improvement in the statistical power of model
selection procedures can improve selection accuracy.

Our results show that, consistent with the findings of \textcite{cooperCrossvalidatoryModelSelection2024},
joint scores outperform univariate scores when correlation between
neighboring points is strong. Our analysis concludes with applications
to real data, with results that are consistent with our simulation
study.

To summarize, in this paper we demonstrate that for several spatial models with
Gaussian covariance structures:
\begin{itemize}
\item Consistent with existing literature, when (strong) spatial effects
are present in candidate models, block CV procedures have lower variance
and better model selection performance than leave-one-out CV (LOO-CV) variants;
\item Moreover, jointly evaluated scoring rules deliver more accurate model
selection outcomes with test statistics that exhibit lower variability
than pointwise alternatives;
\item In Section~\ref{sec:examples}, we present applied examples with
real data.
\end{itemize}
The remainder of this paper proceeds as follows. In Section~\ref{sec:SpatialCV}
we provide a brief overview of spatial CV methods and summarize
related work. In Section~\ref{sec:Simulations} we present simulation
evidence that jointly-evaluated scoring rules outperform pointwise
methods for model selection. In Section~\ref{sec:examples} we apply
the methods to actual data, and Section~\ref{sec:Conclusion} concludes.

\section{Spatial cross-validation\label{sec:SpatialCV}}

Suppose we observe a fixed data vector $y=\left(y_{1},\dots,y_{n}\right)$,
which we will presume drawn from some true (but unknown) spatial process
$p_{\mathsf{true}}\left(y\right)$. The elements of $y$ are indexed
by spatial points or small areas, and $p_{\mathsf{true}}$ embeds
some correlation structure that reflects the spatial relationships
between the elements of $y$. We will use the generic notation $p\left(\cdot\right)$
to denote the density of arguments to the left of any conditioning bar.

Let $\tilde{y}$ denote unseen random values. In spatial modeling
applications, the vectors $y$ and $\tilde{y}$ may or may not overlap
in terms of the geographical regions they index. Model construction begins with the choice of a parametric model family $p\left(y\,|\,\theta\right)$,
where the $\theta\in\Theta$ has finite dimension. Usually, $p_{\mathsf{true}}\left(y\right)$
will not be a member of the posited model family, that is, the model
will be at least somewhat misspecified.

When prior information $p\left(\theta\right)$ about the likely values
of $\theta$ is available, Bayes' rule summarizes the available information
as the posterior density $p\left(\theta\,|\,y\right)\propto p\left(\theta\right)p\left(y\,|\,\theta\right)$,
and from that the predictive density $p\left(\tilde{y}\,|\,y\right)=\int p\left(\theta\,|\,y\right)p\left(\tilde{y}\,|\,\theta\right)\,\mathrm{d}\theta$.

But how should the model family
$p\left(y\,|\,\theta\right)$ be chosen from several plausible candidates? While
theoretical facts about the application may suggest certain choices
over others \emph{a priori}, modern Bayesian analysis workflows \parencite[e.g.][]{gelman2020bayesian}
propose data-driven methods for selecting the best-performing models.
Among these methods, predictive assessment \parencite{gelmanPosteriorPredictiveAssessment1996}
measures the performance of a particular model, say $M$, by the predictive performance
of $p\left(\tilde{y}\,|\,y,M\right)$, where we now introduce the model
as part of the notation.

A conceptual approach to predictive assessment is \emph{external validation}
\parencite{gelmanBayesianDataAnalysis2014}. Suppose for a moment
that $p_{\mathsf{true}}\left(y\right)$ were known to the analyst.
Suppose also that we have a \emph{scoring rule} \parencite{gneitingStrictlyProperScoring2007}\emph{,
}a functional $S\left(f,\tilde{y}\right)$ that numerically assesses
the quality of a predictive distribution $f$ given an actual realization
$\tilde{y}$. While problem-specific scoring rules would be ideal,
a natural summary of the performance of $p\left(\tilde{y}\,|\,y\right)$
is the model $M$ expected score, or \emph{expected log predictive
density,}
\begin{equation}
\mathrm{elpd}\left(M\,|\,y\right)=\int \log p\left(\tilde{y}\,|\,y,M\right) p_{\mathsf{true}}\left(\tilde{y}\right)\,\mathrm{d}\tilde{y}.\label{eq:def:extv}
\end{equation}
If we could compute it, this expression would allow model selection
among a finite set of candidates by simply choosing the model that
achieves the highest score (we use positively-oriented scoring rules).
In the notation, we have placed $y$ on the right-hand-side of the
conditioning bar to emphasize that \eqref{eq:def:extv} is a function
of $y$ via the posterior predictive density $p\left(\tilde{y}\,|\,y, M\right)$.

However, in practice it is rare for $p_{\mathsf{true}}\left(\tilde{y}\right)$
to be known or for independent draws of the data to be available.
We require a feasible alternative to \eqref{eq:def:extv}. Unfortunately,
simply evaluating the predictive score using the training data, i.e.
using $p\left(y\,|\,y, M\right)$ as an estimate for \eqref{eq:def:extv},
would yield optimistically-biased evaluations favoring models that
overfit the data \parencite{vehtariSurveyBayesianPredictive2012}.

CV approximates $\mathrm{elpd}\left(M\,|\,y\right)$ up to a multiplicative
constant using a data-splitting approach. It constructs a Monte Carlo
estimate for \eqref{eq:def:extv} using $K$ divisions of the data
into disjoint training and testing sets. We respectively denote training
and test sets as $y_{\mathsf{train}_{k}}$ and $y_{\mathsf{test}_{k}}$,
for $k=1,\dots,K$. The (jointly-evaluated) CV estimate is given by
\begin{equation}
\widehat{\mathrm{elpd}}_{CV}\left(M\,|\,y\right)=\sum_{k=1}^{K}\log p\left(y_{\mathsf{test}_{k}}\,|\,y_{\mathsf{train}_{k}},M\right),\label{eq:def:cv}
\end{equation}
where $p\left(y_{\mathsf{test}_{k}}\,|\,y_{\mathsf{train}_{k}},M\right)$
denotes a joint predictive density given the $k$th training set,
evaluated at the $k$th test set. An alternative, pointwise-evaluated
formulation is given by
\begin{equation}
\widehat{\mathrm{elpd}}_{CV}^{pw}\left(M\,|\,y\right)=\sum_{k=1}^{K}\sum_{i=1}^{n_{k}}\log p\left(y_{\mathsf{test}_{k},i}\,|\,y_{\mathsf{train}_{k}},M\right),\label{eq:def:cv-pw}
\end{equation}
where $n_{k}$ denotes the size of the $k$th test set, and $p\left(y_{\mathsf{test}_{k},i}\,|\,y_{\mathsf{train}_{k}},M\right)$
represents a univariate predictive density evaluated at the $i$th element of the test set. The CV estimate $\widehat{\mathrm{elpd}}_{CV}\left(M\,|\,y\right)$
is an estimate of model predictive ability. In addition, the score
difference between model $M_A$ and model $M_B$ (which could be computed either pointwise or jointly),
\begin{equation}
\widehat{\mathrm{elpd}}_{CV}\left(M_{A},M_{B}\,|\,y\right)=\widehat{\mathrm{elpd}}_{CV}\left(M_{A}\,|\,y\right)-\widehat{\mathrm{elpd}}_{CV}\left(M_{B}\,|\,y\right),\label{eq:pairwise}
\end{equation}
can be interpreted as a pairwise model selection statistic for selecting
between model $M_{A}$ and model $M_{B}$, given data $y$. Since
$\widehat{\mathrm{elpd}}_{CV}\left(M_{A},M_{B}\,|\,y\right)$ is a
statistic of $y$ it is subject to sampling variability, and its frequency
properties should be considered when using $\widehat{\mathrm{elpd}}_{CV}\left(M_{A},M_{B}\,|\,y\right)$
as the basis for model selection decisions \parencite{sivulaUncertaintyBayesianLeaveoneout2022}.
One popular approach to conducting inference for $\widehat{\mathrm{elpd}}_{CV}\left(M_{A},M_{B}\,|\,y\right)$
is to assume that the (large number of) contributions in \eqref{eq:def:cv}
are independent and have finite variance, so that a Gaussian approximation
for the distribution of $\widehat{\mathrm{elpd}}_{CV}\left(M_{A},M_{B}\,|\,y\right)$
is appropriate \parencite{vehtariBayesianModelAssessment2002,sivulaUncertaintyBayesianLeaveoneout2022}.

Aside from CV, a huge range of alternative model selection techniques
is available, many of which apply to Bayesian spatial modeling
problems \parencite{murModelSelectionStrategies2009}. These include
marginal likelihood methods \parencite{bernardoBayesianTheory2000}
and information criteria \parencite{gelmanUnderstandingPredictiveInformation2014,vehtariPracticalBayesianModel2017}.
CV is attractive for several reasons. It is extremely general and
often straightforward to implement. In addition, CV avoids sensitivities
to prior specification inherent in marginal likelihood methods \parencite{lindley1957statistical}
and is appropriate in `$\mathcal{M}$-open' settings \parencite{bernardoBayesianTheory2000}
where the model is known to be at least somewhat misspecified, i.e.
where $p_{\mathsf{true}}\not\in\left\{ p\left(\cdot\,|\,\theta\right)\,:\,\theta\in\Theta\right\} $,
which is the case in most applications in economics, ecology, and
policy analysis, among others \parencite{kelterBayesianModelSelection2021}.

For models of non-\emph{iid} data, such as those with spatial effects,
CV needs to be implemented carefully to ensure that the summands in
the estimator \eqref{eq:def:cv} are close to independent \parencite{hastie2009elements,arlotSurveyCrossvalidationProcedures2010}.
Independence of the summands is trivially satisfied in the special
case where the observations $\left(y_{i}\right)$ are \emph{iid},
but not when the $\left(y_{i}\right)$ exhibit more general structures
like temporal or spatial dependence. Where residual spatial autocorrelation remains between CV
folds, CV is likely to over-estimate predictive performance \parencite{telfordEvaluationTransferFunctions2009,lerestSpatialLeaveoneoutCrossvalidation2014,pohjankukkaEstimatingPredictionPerformance2017,robertsCrossvalidationStrategiesData2017,deppnerAccountingSpatialAutocorrelation2022}.
Specialized spatial cross-validation methods are therefore helpful.

Broadly speaking, standard CV methods are adapted to spatial settings
by the removal of a `buffer' or `halo' around the test set to
ensure near independence of $y_{\mathsf{train}}$ and $y_{\mathsf{test}}$
(see e.g. \cite{lerestSpatialLeaveoneoutCrossvalidation2014,pohjankukkaEstimatingPredictionPerformance2017,beigaiteSpatialCrossValidationGlobally2022}),
the use of `blocks' of contiguous data points for test sets (e.g. \cite{pohjankukkaEstimatingPredictionPerformance2017,robertsCrossvalidationStrategiesData2017,mahoneyAssessingPerformanceSpatial2023}),
and avoiding completely random blocking methods that ignore the spatial
structure \parencite{wengerAssessingTransferabilityEcological2012}.
Some approaches adopt data- or model-specific information to choose
the folds (e.g. \cite{liuLeavegroupoutCrossvalidationLatent2023}).
While specific structures are not our focus, we will adopt `block
CV' as described by \textcite{robertsCrossvalidationStrategiesData2017}
as representative of common methods. Block CV chooses contiguous blocks
of test observations. The test set is possibly separated from the
training set by an additional halo of observations removed from the
training set to ensure $y_{train}$ and $y_{test}$ are close to independent.
\textcite{mahoneyAssessingPerformanceSpatial2023} analyze several flavors
of blocked CV, and in results that appear to broadly agree with \textcite{robertsCrossvalidationStrategiesData2017}
find some advantage to constructing blocks in a data-driven manner
rather than by dividing the available space by tiling.

One aspect of spatial CV that remains application-specific is the
need for `spatial stratification'. Constructing appropriate spatial
CV methods is complicated by the difficulty of making accurate predictions
in unobserved geographic areas, which forces the model to `interpolate'
rather than predict within its training range \parencite{adinAutomaticCrossvalidationStructured2023}.
Data-driven stratification can lead to bias \parencite{karasiakSpatialDependenceTraining2022}.
Broadly speaking, model generalizability depends on model specification
and complexity, as well as the characteristics of the data generating
process under study \parencite{lieskeRobustTestSpatial2011}.

In contrast to the benefit of blocked test sets \parencite{robertsCrossvalidationStrategiesData2017},
scoring approaches have received less attention. The scoring approach
incorporates the scoring rule in use and whether it is multivariate
(for joint evaluation, as in \eqref{eq:def:cv}) or univariate (pointwise,
as in \eqref{eq:def:cv-pw}). Popular objective functions include
the area under the receiver operating characteristic curve \parencite[AUC;][]{hastie2009elements}
(see e.g. \cite{wengerAssessingTransferabilityEcological2012,brenningSpatialCrossvalidationBootstrap2012})
for categorical models and root mean square error \parencite[RMSE;][]{hastie2009elements}
for continuous response variables. In models capable of probabilistic
predictions, the pointwise elpd \parencite{vehtariPracticalBayesianModel2017},
also known as the conditional predictive ordinate \parencite[CPO;][]{adinAutomaticCrossvalidationStructured2023},
is most often deployed. However, the examples referenced here are
computed pointwise: they aggregate contributions from individual predicted
points within the test set, as in \eqref{eq:def:cv-pw}.

Our focus in this paper is areal spatial models, where for some small
area~$t$, $y(t)$ follows some observation density $p(y(t)\,|\,\lambda\left(z(t)\right),\theta)$
with link function $\lambda\left(\cdot\right)$ and latent values $z(t)$ that can be decomposed as
\begin{equation}
z\left(t\right)=\beta^{\top}x\left(t\right)+f\left(t\right)+\varepsilon\left(t\right),\label{eq:spatial-gen-form}
\end{equation}
where $\beta$ denotes a regression coefficient, $x\left(t\right)$
is a vector of spatially-indexed explanatory variables, $f\left(t\right)$
is a spatial effect, and $\varepsilon\left(t\right)$ is an individual
effect. A common example of \eqref{eq:spatial-gen-form} is the class
of Gaussian Markov random field models (GRMFs; \cite{bivandApproximateBayesianInference2014,liuLeavegroupoutCrossvalidationLatent2023}),
where $f\left(t\right)$ has a Gaussian covariance with sparse precision.
The GRMF class includes as special cases simultaneous autoregressive
SARs and CARs.
A range of alternatives are available for various applications \parencite{anselin1988spatial}.
Hierarchical models \parencite{banerjeeHierarchicalModelingAnalysis2015},
formulations popular for disease mapping applications \parencite{rieblerIntuitiveBayesianSpatial2016,lerouxEstimationDiseaseRates2000},
and other more specialized formulations \parencite{utaziBayesianLatentProcess2018}.
In many applications, where data limitations or a lack of theory lead
to uncertainty about the appropriate form of candidate models, high-capacity
machine learning models such as random forests that do not directly
account spatial dependence are popular \parencite[e.g.][]{lerestSpatialLeaveoneoutCrossvalidation2014,robertsCrossvalidationStrategiesData2017}.
In these cases, the lack of an explicit covariance function for $f\left(t\right)$
means that it is difficult to construct the joint density $p\left(y_{\mathsf{test}_{k}}\,|\,y_{\mathsf{train}_{k}},M\right)$
that appears in \eqref{eq:def:cv}.

A particular challenge facing CV for spatial applications is the high
computational cost of each model fit. The main problem is that naively
computing covariance functions usually requires inverting and/or computing
log determinants of $n\times n$ matrices, which in general requires
$O\left(n^{3}\right)$ floating-point operations or \emph{flops} \parencite{simpson2012order}.
In general, this cost is multiplied by the number of CV folds, and
usually at least by the number of iterations of an inference algorithm,
which can be especially costly for Monte Carlo Markov (MCMC) chain
inference.

Although reducing the computational cost of spatial inference and
CV is not the focus of this paper, it is worth noting that faster
approximate inference is available for special cases. For instance,
tractable MCMC samplers are available for SAR \parencite{lesageBayesianEstimationSpatial1997}
and CAR models \parencite{doneganCARModelsStan2022}. Approximate
methods are available when datasets are large \parencite{burdenSARModelVery2015,zhangKrigingCrossvalidationMassive2010}.
\textcite{lindgrenExplicitLinkGaussian2011} reformulate GMRFs as
the solution to stochastic partial differential equations, leading
to cheaper approximate computation methods. Notably, this method is
implemented as part of the R-INLA software suite \parencite{gomez-rubioSpatialModelsUsing2019}.
\textcite{liuLeavegroupoutCrossvalidationLatent2023} further approximate
leave-group-out-CV using R-INLA, by extending \textcite{heldPosteriorCrossvalidatoryPredictive2010} and
\textcite{vehtari2016bayesian}.
Other approximate CV approaches for specific models include \textcite{wood2024neighbourhood},
who approximates a CV estimator for a quadratic loss surface using
Newton update steps.

\section{Simulation study\label{sec:Simulations}}

In this section we investigate joint spatial cross-validation for model
selection with three simulation studies. The selection exercises
demonstrate selection of the regression parameter, spatial network weights,
and model form. The goal is to compare pointwise versus joint evaluation, 
and also to demonstrate the impact of departing from LOO-CV by reducing
the number of CV folds and increasing test size. The primary measure of
performance here is the probability of correct model selection.

We begin with model selection experiments using SARs (Subsection~\ref{subsec:sar}), which are widely used
in spatial data analysis . The SAR model is mathematically
very similar to the CAR model, indeed in a certain sense they are
equivalent \parencite{verhoefRelationshipConditionalCAR2018}.
For simplicity of exposition, each experiment demonstrates several
pairwise comparisons, presented as pairwise model comparison statistics
defined in \eqref{eq:pairwise}. For simplicity, in each comparison positive
values indicate the choice of the better model and vice-versa.

We are interested in the frequency properties of $\widehat{\mathrm{elpd}}_{CV}\left(M_{A},M_{B}\,|\,y\right)$
under different cross-validation approaches, across many different
independent realizations of the data vector $y\sim p_{\mathsf{true}}$, following \textcite{sivulaUncertaintyBayesianLeaveoneout2022}
and \textcite{cooperCrossvalidatoryModelSelection2024}. The resulting distribution of selection statistics is nonstandard \parencite{bengioNoUnbiasedEstimator2004a}. 
We are interested in the share of this distribution that falls to the left or right of 
zero, so is useful to characterize this distribution by the ratio of its mean to its 
standard deviation,
\begin{equation}
Z=\frac{\frac{1}{N}\sum_{i=1}^N \widehat{\mathrm{elpd}}_{CV}\left(M_A, M_B\,|\,y_i\right)}{\sqrt{\mathrm{var}_i\left(\widehat{\mathrm{elpd}}_{CV}\left(M_A, M_B\,|\,y_i\right)\right)}}.
\label{eq:def:pop-zratio}
\end{equation}
In \eqref{eq:def:pop-zratio}, $N$ is the number of independent experiment replications 
(noted below for each experiment) and 
$\mbox{var}_i\left(\cdot\right)$ in the denominator is the sample
variance across independent replication draws.
We will refer to \eqref{eq:def:pop-zratio} as the $Z$ ratio, in a nod to 
its similarity to a hypothesis test for the sign of $\mss$.

Bayesian CV applications are computationally expensive, repeated applications for
multiple independent replications and sequences of experiments even more so. 
Accordingly, the main
practical challenge to be overcome for this simulation study is to reduce the inference
costs of repeated model fits enough for the study to be computationally feasible. To do so, we will
adopt two shortcuts. First, we use models with Gaussian observation densities
and linear link functions. This allows $z$
to be integrated out analytically (see e.g., \cite{banerjee2020modeling}, \S1).
Second, we use Laplace approximation for
inference \parencite{mackay2003information}.
While Laplace approximation admittedly
can be a crude posterior approximation and hence CV score estimate,
computations are relatively cheap and easily vectorized, and 
the resulting approximate posteriors appear to be very similar to more accurate
MCMC-based estimates. See Appendix~D for details.

\begin{figure}
\begin{centering}
\includegraphics[width=1\textwidth]{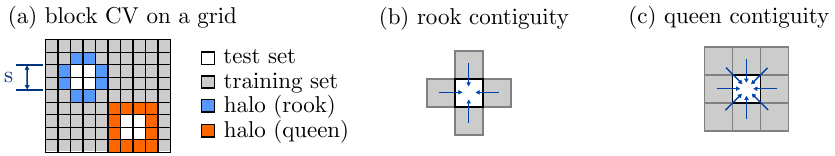}
\par\end{centering}
\caption{Cross-validation grid design used for simulation experiments. Panel
(a) shows a rectangular grid with rectangular test sets of side $s$
(so $n_{\mathrm{test}}=s^{2}$), separated by a degree-1 halo. The
shape of halo depends on the assumed contiguity relationship (rook
or queen). Panel (b) shows dependence relationships (blue arrows)
for a generic grid square (white) and its contiguous neighbors (gray)
under rook contiguity. Panel (c) shows the same under queen contiguity.}
\label{fig:block-cv}
\end{figure}

The three sequences of experiments in Sections~3.1.1, 3.1.2, and 3.2 are each 
conducted on a square regular lattice (Figure~\ref{fig:block-cv}),
with square test sets chosen to completely tile the plane. 
The regular lattice has $n=576$,
with rook contiguity unless otherwise noted. This means that the 
number of CV folds decreases from $K=576$ (for $1\times 1$ test sets)
as the test set size grows. For example, for $4\times4$ test sets
we have $K=576/16=36$. Within each experiment sequence, the $N$ independent generated datasets are common for all test set sizes.
For consistency, an order-1 halo is used throughout.
The number of independent replications $N$ differs across experiments,
reflecting varying computational cost.

\subsection{Simultaneous autoregressive model (SAR)\label{subsec:sar}}

SAR models are a popular workhorses of economic and ecological analysis
\parencite{cressie2015statistics}. The spatial regressions we are
interested in have the form
\begin{equation}
\left(I_{n}-\rho W_{+}\right)y=X\beta+\varepsilon,\label{eq:spatialreg}
\end{equation}
where $X$ is an $n\times k$ matrix of explanatory variables measured
at the nodes, and $\varepsilon\sim\mathcal{N}\left(0,\sigma_{\varepsilon}^{2}I\right)$
is an \emph{iid} noise vector.

We standardize $\rho$ to the unit interval, which allows us to compare
the strength of dependence from one model to another. Following \textcite{verhoefSpatialAutoregressiveModels2018},
let $W$ denote the adjacency matrix for the undirected spatial network
with $n$ nodes. In typical applications of SAR models $W$ is sparse,
so that most elements are zero. A number of neighborhood structures
are available, perhaps the simplest being contiguity, where
\begin{equation}
W_{ij}:=\begin{cases}
1 & \textnormal{if \ensuremath{i} and \ensuremath{j} are connected},\\
0 & \textnormal{otherwise.}
\end{cases}
\end{equation}
We follow \textcite{verhoefSpatialAutoregressiveModels2018}
and row-standardize the weights matrix by defining 
\begin{equation}
\left(W_{+}\right)_{ij}:=W_{ij}/\left(\sum_{i}W_{ij}\right),
\end{equation}
so that $\sum_{i}\left(W_{+}\right)_{ij}=1$. Row-standardization ensures that the
spatial precision matrix $\Sigma^{-1}=\left(I_{n}-\rho W_{+}\right)$
is guaranteed to exist and be positive definite for all $\rho\in\left[0,1\right)$.

In this sequence of experiments, we perform separate model selection
procedures to simulate selecting three aspects of the model: the covariates
$X$, the underlying network (i.e. the choice of $W$); the order
of the lag structure. For each case, we vary (a) the size of the test
set and (b) the value of the dgp's autoregressive parameter, which
controls the degree of persistence of the data. The covariate matrix
$X$ contains a constant column of 1s and the remaining columns are
independent standard normal draws. $X$ is re-drawn independently
each repetition, so that the results smooth over randomness in $X$.
Throughout we impose a variety of arbitrary, but plausible, weakly-informative
priors.

\begin{figure}
\begin{centering}
\includegraphics[width=1\textwidth]{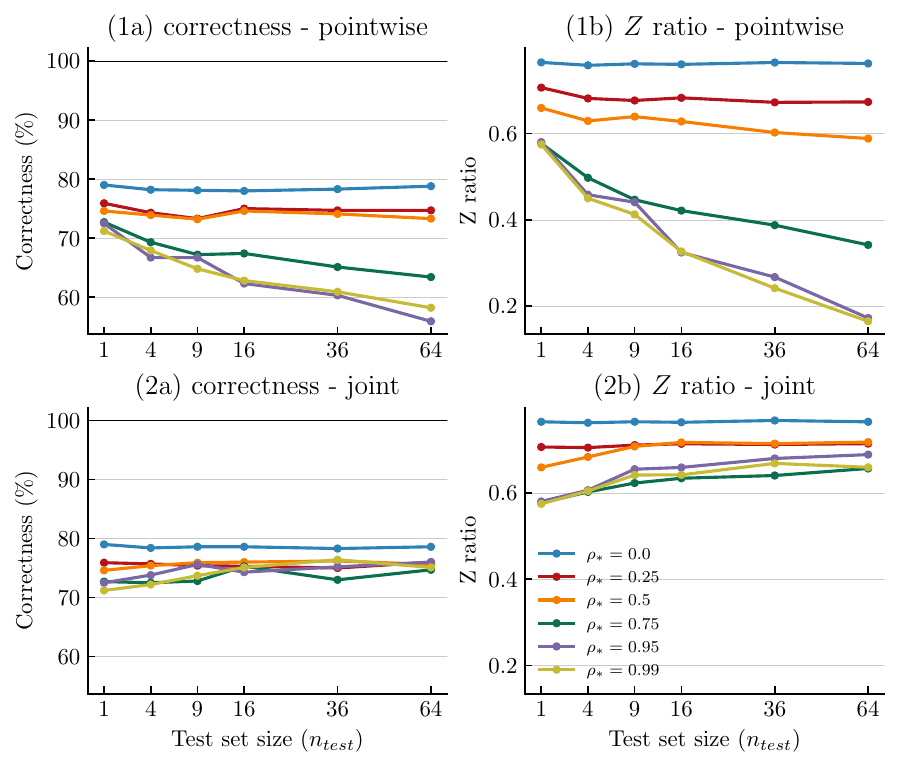}
\par\end{centering}
\caption{
Summary of a sequence of model selection experiments to 
identify model regression component, each with 1,000 independent replications, for 
SAR models on a regular 
lattice with $n=576$ (see Section~\ref{subsec:SAR-X}). Model selection is by blockwise CV under the logarithmic score with the block
size on the $x$-axis. The model and dgps are Gaussian SARs, with
$\left(I_n - \rho W_+\right)y=X\beta + \sigma\varepsilon$.
True $\sigma_*^2=5$, $\beta_{*}=\left(1,1,0.9\right)$,
and $X$ is a $n\times3$ matrix containing a column of ones and 
standard normal draws, and $W_+$ reflects rook adjacency.
Covariates are
common to dgp and both candidates. Several true autoregressive parameters $\rho_*$
are plotted, indicated by color. $M_{A}$ is missing the third covariate and $M_{B}$ lacks
the second; correct model selections identify the $M_{A}$. Panels (1a) and (2a) plot the share
of independent data draws where the correct model is selected. Panels
(1b) and (2b) plot the $Z$ ratio of the resulting distribution
of model selection statistics $\mss$, oriented so that positive values
indicate correct model selection. Panels (1a) and (1b) are computed using
pointwise evaluation; (2a) and (2b) joint evaluation. Joint evaluation yields higher
accuracy and lower relative variability, as indicated by $Z$ ratio.
}
\label{fig:sar-beta-sel}
\end{figure}

\subsubsection{Covariate selection\label{subsec:SAR-X}}

We begin with a sequence of experiments focusing on the selection of covariates in the regression
component of the SAR model. Both candidate models are misspecified,
since neither $M_{A}$ nor $M_{B}$ has the same form as dgp. The
underlying dgp has true parameter $\beta_{0}=\left(1,1,0.9\right)$.
Candidate $M_{A}$ has access to the first two elements of $X$ and
$M_{B}$ has the first and third. Since the explanatory power of $M_{A}$
is larger, cross-validation should tend to prefer $M_{A}$. We perform $1,000$
repetitions of the model selection process, for each of $\rho_{*}\in\left\{ 0.0,0.25,0.5,0.75,0.95,0.99\right\} $
and square test set sides $s\in\left\{ 1,2,3,4,6,8\right\} $, so
that $n_{\mathsf{test}}=s^{2}$. In each case, the noise variance
$\sigma_{*}^{2}=5$. These dgp parameter values were chosen to make
the selection process difficult but not too difficult. If $\left|\beta_{0}\right|$
were much larger, say, then selection would be too easy and CV would
almost always identify the correct candidate, in this case $M_{A}$.
Priors are $\beta\sim N\left(0,I_{k}\cdot10\right)$, $\rho\sim\mathrm{Beta}\left(2,2\right)$,
and $\sigma^{2}\sim\mathcal{N}_{+}\left(0,10\right)$, where $\mathcal{N}_{+}$
denotes the positive half-normal distribution. 

Figure~\ref{fig:sar-beta-sel} summarizes the results for each $\rho_{*}$
value, as the test set size increases. The results are consistent
with those presented by \textcite{cooperCrossvalidatoryModelSelection2024}
for univariate autoregressions. The results differ according to the
strength of the dependence parameter $\rho_{*}$. Under weak dependence
(for $\rho_{*}\leq0.5$ in this experiment), size of the test set
and evaluation method do not influence selection accuracy much. In
contrast, when dependence is stronger ($\rho_{*}\geq0.75$), test set
size and evaluation method strongly influence selection accuracy.
Under pointwise evaluation of the logarithmic score, accuracy declines
with increasing test set size, an effect that is stronger for larger
values of $\rho_{*}$ (Panel~(1a)). When the scoring rule is evaluated
jointly, accuracy shows moderate improvements with larger test set
sizes, especially for larger $\rho_{*}$ values (Panel~(2a)).

Taken together, the results suggest that under the logarithmic score,
joint evaluation develops greater statistical power for covariate
selection than pointwise evaluation. Under stronger correlation
(greater $\rho_*$), the difference is larger
when test sets include greater numbers of
predictions, so that test sets are better able to fully capture the correlation
structure of the data (\cite{cooperCrossvalidatoryModelSelection2024}). Changes in model selection power
are explained by shifts in both the location and variability of the
distribution of $\mss$ across $y$ draws. Panels~(1b)
and (1c) succinctly summarize these changes by the Z ratio,
$\mathrm{Z}:=\hat{\mu}/\hat{\sigma}$,
for $\hat{\sigma}$ and $\hat{\mu}$ respectively
the sample standard error and mean of $\mss$, across all
$y$ draws. Although both location and variability of this distribution
influence model selection accuracy, $Z$ summarizes the overall
impact on accuracy. For an alternative graphical summary of these
distributional shifts, see Figure~B.1 in Appendix~B.

\subsubsection{Network structure selection\label{subsec:SAR-W}}

\begin{figure}
\begin{centering}
\includegraphics[width=1\textwidth]{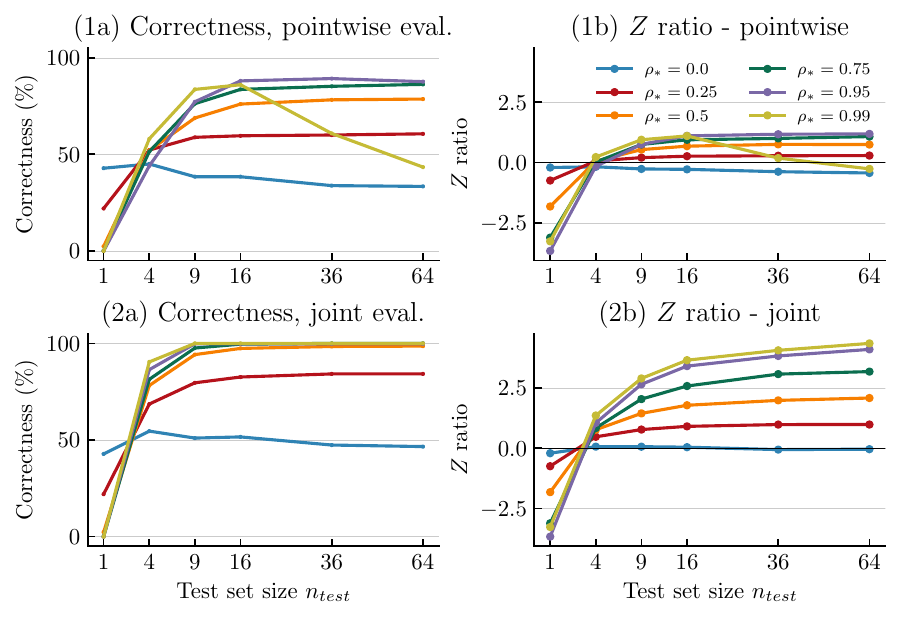}
\par\end{centering}
\caption{Summary of a sequence of model selection experiments to 
identify the model network, each with 500 independent replications, for 
SAR models a regular 
lattice with $n=576$. Model selection is by blockwise CV with the block
size noted on the $x$-axis. The model and dgps are Gaussian SARs, with
$\left(I_n - \rho W_+\right)y=X\beta + \sigma\varepsilon$.
True $\sigma_*^2=4$, $\beta_*=(1,1,1)$ and $X$ 
is a $n\times3$ matrix containing a column of ones and 
standard normal draws. Candidate models are distinguished by $W_+$,
the row-standardized agency matrix.
The dgp has rook adjacency and candidate models include rook
and queen adjacency (Figure~1). Covariates are
common to dgp and both candidates. Colors indicate the true $\rho_*$
parameter. Panels (1a) and (2a) plot the share
of independent data draws where the correct model is selected. Panels
(1b) and (2b) plot the $Z$ ratio of the resulting distribution
of model selection statistics $\mss$, oriented so that positive values
denote correct model selection. Panels (1a) and (1b) are computed using
pointwise evaluation; (2a) and (2b) joint evaluation. With the exception 
of the case where $\rho_*=0$, joint evaluation yields higher
accuracy and lower relative variability ($Z$ ratio).
See also Figure~B.2 in Appendix~B.1.
}

\label{lbl:sar-connectivity-sel}
\end{figure}
In SAR models the spatial structure is encoded in the weight matrix,
which can be constructed a variety of ways such as discrete
adjacency, distance-based methods, and kernel methods. In this sequence
of experiments we select between two subtly different structures:
discrete rook and queen adjacency for the same regular lattice (Figure~\ref{fig:block-cv}).
The dgp has rook adjacency. The two candidate models are a SAR with
rook and queen adjacency, and covariates are fully observed and common
to both models. That is, one candidate model is correctly specified.
Positive values of $\widehat{\mathrm{elpd}}_{CV}\left(M_{A},M_{B}\,|\,y\right)$
indicate correct model selection. 

Figure~\ref{lbl:sar-connectivity-sel} summarizes the results of
500 independent model selections across independent data draws. When
the underlying dgp has no autoregressive dependency at all ($\rho_{*}=0$),
the selection experiment is under-determined and CV is unable to select
between the candidate models, with success rates around 50~per~cent.
Of course, when $\rho_{*}=0$, all observations are independent and
hence there is no signal in the data that would indicate the true
underlying dependence structure.

When dependence is present ($\rho_{*}>0$), selection correctness
improves for larger $n_{\mathsf{test}}$ under joint evaluation (Panel
(2a)). This is also seen for pointwise evaluation, although overall
selection performance is worse than under joint evaluation. Stronger
dependence in the underlying data generally results in better selection
performance, except under very strong dependence $\left(\rho_{*}=0.99\right)$
selection performance falls off quickly under pointwise evaluation.
As in the previous experiment, these trends are explained by the relationship
between the location and variability of the distribution of $\widehat{\mathrm{elpd}}_{CV}\left(M_{A},M_{B}\,|\,y\right)$.
Under joint evaluation, variability increases less compared to shifts
in the overall distribution, resulting in greater model selection
power (Panels (1b) and (2b)). See Figure~B.2
in Appendix~B.1 for an alternative graphical
description of the distribution of model selection statistics.
Also see Appendix~A for alternative experiments
where the dgp has queen adjacency, and where the neighborhood is distinguished
by different orders (numbers of steps).

\subsection{Covariance kernel selection\label{subsec:latgauss}}

\begin{figure}
\begin{centering}
\includegraphics[width=1\textwidth]{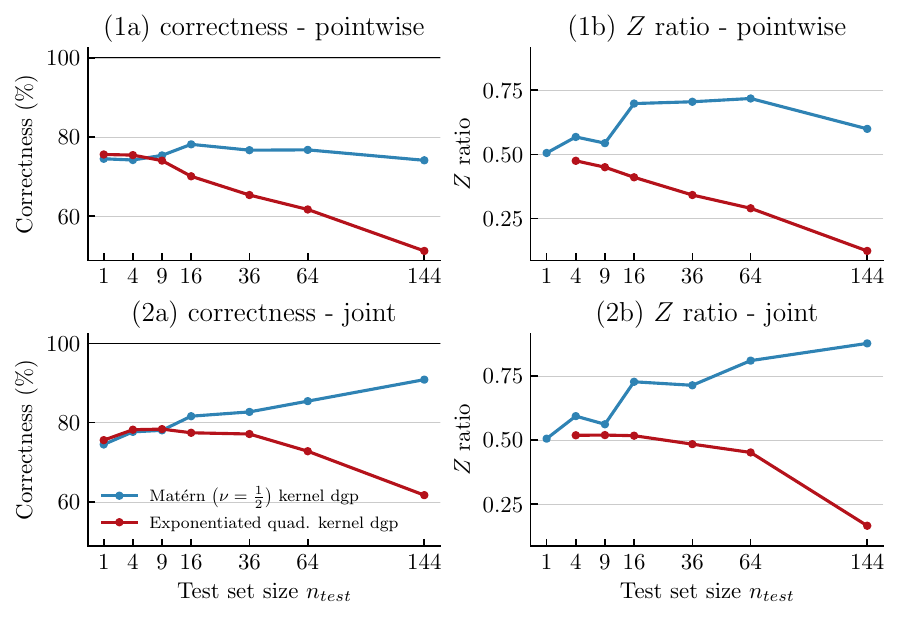}
\par\end{centering}
\caption{Summary of a sequence of experiments selecting between two
models with Gaussian covariance structures, for different test set sizes
(Section~\ref{subsec:latgauss}).
Each experiment is performed independently 1,200 times with independent
data draws. Experiment sequences with two different dgps are shown: a Mat\'ern
kernel with $\nu=\frac{1}{2}$ ({\color{blue}blue line}), and exponentiated quadratic
kernel ({\color{red}red line}). True dgp length scale and noise variance are set to 1.
Candidate models have (i) Mat\'ern kernel with $\nu=\frac{1}{2}$
and (ii) an exponentiated quadratic kernel, but with random (estimated) parameters.
$M_A$ is the correct model and vice-versa. For both dgps, CV with joint evaluation (Panel (1a)) outperforms pointwise-evaluated
methods CV (Panel (1b)) in terms of correctness.
Broadly speaking this is associated with lower relative
variability/higher Z-scores for the joint methods (Panels (1b) and (2b)).
In Panels (1b) and (2b) we reduce noise by applying a 98\% trimmed mean and variance.
}

\label{lbl:matern}
\end{figure}

To show that the previous example is not unique to SAR structure,
we now repeat the analysis with model with an alternative, but still
Gaussian, covariance structure. In this case we use CV to
select between models defined by covariance kernels, again for simulated data
on a regular lattice. 

For simplicity, we omit regression components, so models we consider
here are have the form $p\left(y\,|\,\theta\right)\sim N\left(y\,|\,0,K_{\theta}\right)$.
Unlike the SAR which has a sparse covariance, here $K_{\theta}$ is
dense, defined by a covariance kernel applied to the Euclidean $d$
distance between area midpoints. This example is motivated by the
fact that in applied settings, the appropriate choice of kernel functions
for $K_{\theta}$ is seldom clear and CV is often used make the selection
\parencite{arlotSurveyCrossvalidationProcedures2010}.

We run two sequences of experiments, each with different dgps. Both dgps
have isotropic covariance kernels: the first has a Mat\'ern kernel with $\nu=\frac{1}{2}$,
given by $K\left(d\right):=\sigma^{2}\exp\left(-d/\lambda\right)$,
and the second an exponentiated quadratic kernel, defined as $K\left(d\right):=\sigma^{2}\exp\left(-d^{2}/\left(2\lambda^{2}\right)\right)$.
Both fix true parameters $\lambda_{*}=\sigma_{*}=1$.

For both sequences of experiments, the two candidate models have Mat\'ern and exponentiated quadratic kernels, but with the parameters $\lambda$ and $\sigma$ random (estimated).
To close the model, we impose the priors $p\left(\sigma\right)=\mathcal{N}_{+}\left(\sigma\,|\,0,1\right)$
and $p\left(\lambda\right)=\mathcal{N}_{+}\left(\lambda\,|\,0,1\right)$,
where $\mathcal{N}_{+}$ denotes the positive half-normal distribution.
To ease interpretation, we always choose the candidate $M_A$ so that it matches the true dgp, and $M_B$ the 
incorrect alternative (i.e. score differences are positive
if selection is correct).

The results plotted in Figure~\ref{lbl:matern} are consistent with
two stylized facts evident in previous results, despite the plots being
somewhat noisy. First, model selection
performance at least moderately improves with multivariate test sets,
although in all but one case, selection performance drops off as $n_{test}$
grows as $n_{train}$ decreases. Second, when $n_{test}>1$, jointly-evaluated
scoring rules (Panel (2a)) outperform pointwise evaluated scoring
rules (Panel (1a)). As in earlier experiments, this discrepancy is
explained by a distribution of test statistics that is better separated
from zero, indicated by a $Z$~ratio with larger magnitude (Panels\,(1b)\,and\,(2b)).
For an alternative visualization of changes in the distribution of
$\widehat{\mathrm{elpd}}_{CV}\left(M_{A},M_{B}\,|\,y\right)$ see Figure~B.6 in Appendix~B.2.

Taken together, these simulations show that leave-group-out CV with a dozen or
few dozen folds performs well in
a variety of spatial modeling settings, compared with LOO-CV. When dependence
is strong (in these simulations, standardized $\rho>\frac{3}{4}$), joint CV has
better model selection accuracy than when the logarithmic score is evaluated
pointwise.

\section{Case studies\label{sec:examples}}

In this section we apply CV to models of real data. Naturally, with
real data $p_{\mathrm{true}}$ is not known, so direct measurements
of CV uncertainty by repeated simulation are unavailable. As such,
only a single realization of each pairwise model comparison $\widehat{\mathrm{elpd}}_{CV}\left(M_{A},M_{B}\,|\,y\right)$
is available, conditioned on the observed data vector $y$. Furthermore,
the distribution of $\widehat{\mathrm{elpd}}_{CV}\left(M_{A},M_{B}\,|\,y\right)$
is unobservable. Nonetheless an estimate of the population variance is available by
Gaussian approximation (\cite{sivulaUncertaintyBayesianLeaveoneout2022};
Section~\ref{sec:SpatialCV} of this paper), which rests on the assumption that there are a large number of contributions to 
$\widehat{\mathrm{elpd}}_{CV}\left(M_{A},M_{B}\,|\,y\right)$ that are not strongly correlated,
and that have finite variance.

Under the Gaussian approximation, we have a sample analog to \eqref{eq:def:pop-zratio},
\begin{equation}
\widehat{Z}_y = \frac{\mss}{\sqrt{\frac{K}{K-1}\sum_{k=1}^K\left(\delta_k - \frac1K \sum_{i=1}^K\delta_k \right)^2}},
\label{eq:def:sample-z}
\end{equation}
where $\delta_k = \log p\left(y_{\mathsf{test}_k}\,|\,p_{\mathsf{train}_k}, M_A\right) - \log p\left(y_{\mathsf{test}_k}\,|\,p_{\mathsf{train}_k}, M_B\right)$ is the fold $k$ contribution
to $\mss$. The denominator represents the standard deviation of $\widehat{\mathrm{elpd}}_{CV}\left(M_{A},M_{B}\,|\,y\right)$ under the Gaussian approximation. We refer to \eqref{eq:def:sample-z} as the $\widehat{Z}_y$
ratio.

The applications illustrate the stylized facts presented above: when
spatial autocorrelation is strong (standardized $\rho>\frac{3}{4}$),
joint CV has lower relative variability, measured by $\hat{Z}_y$. However,
when spatial autocorrelation is weaker, this effect is negligible.

As with many ecological public health studies, the focus of both applications
is the small areas themselves rather than the individuals within them.
This is relevant in view of the `ecological fallacy', where relationships
in analyses of aggregated data may not be evident at the individual
level \parencite{cressie2015statistics,gotway2002combining}. Indeed,
the aggregation scheme itself can influence results \parencite{openshaw1984modifiable}.
In a public health context where resources are allocated by geographical
area, the appropriate aggregation scheme is the aggregation scheme
used by the public health authority.

We perform inference using \texttt{R-INLA} version 24.06.27 \parencite{ruehavardINLAFullBayesian2023}
for the \texttt{R} language version 4.4.1 \parencite{rcoreteamLanguageEnvironmentStatistical2024}.
Cross-validation is performed by brute force: the model is fit multiple
times with different data subsets, and scoring is performed by simulating
5,000 draws from the posterior distribution.

\subsection{Child non-vaccination in Australia\label{subsec:child-imm}}

\begin{figure}
\begin{centering}
\includegraphics[width=1\textwidth]{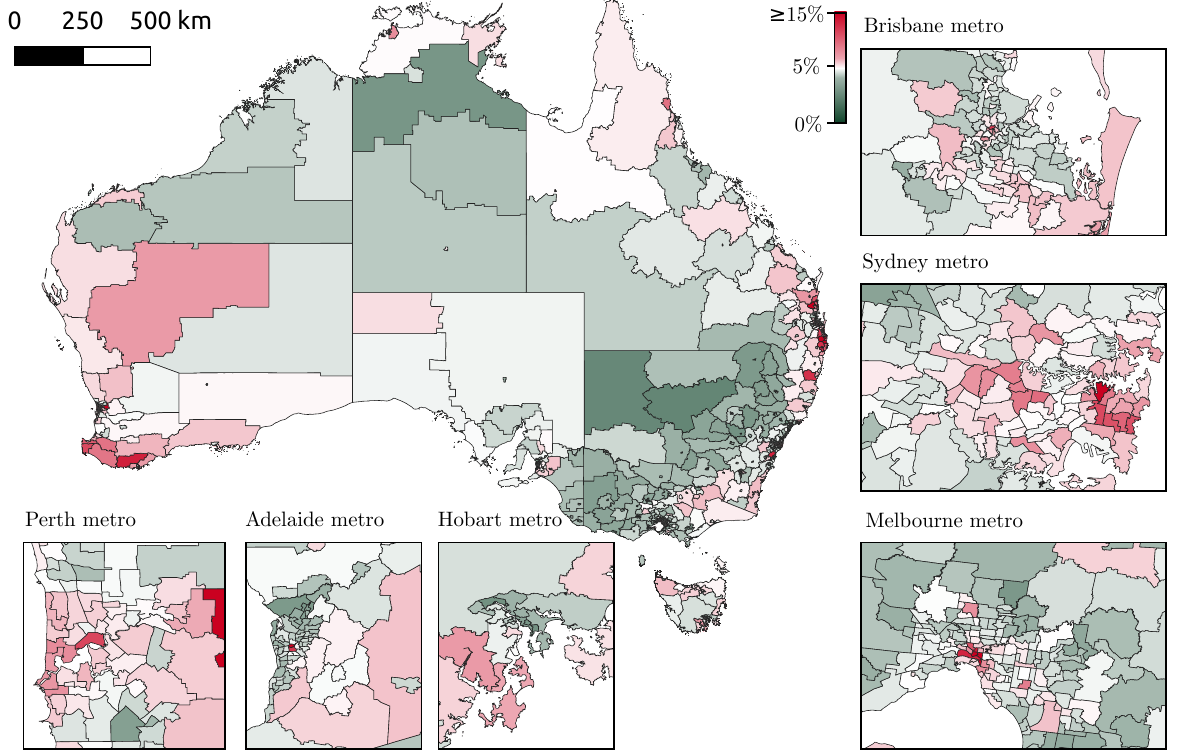}
\par\end{centering}
\caption{A map of Australia, and six metro area regions.
The colors shown relate to fitted values of the preferred model
for the Australian child non-vaccination
rates example (Section~\ref{subsec:child-imm}), reflecting considerable regional
variation and spatial autocorrelation of contiguous areas.}
\label{fig:imm-fitted}
\end{figure}

\begin{figure}
\begin{centering}
\includegraphics[width=1\textwidth]{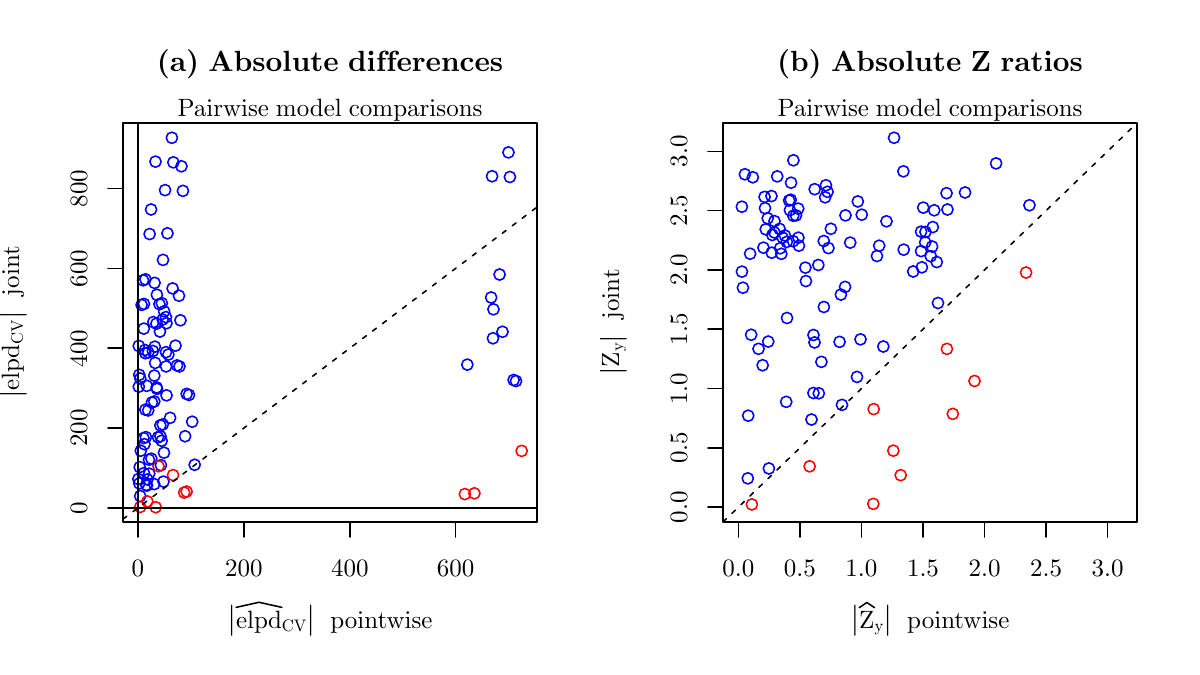}
\par\end{centering}
\caption{Summary of 105 pairwise comparisons among 15 candidate spatial models
for the Australian child non-vaccination rates example (Section~\ref{subsec:child-imm}).
Panel (a) compares the magnitude of pointwise and joint $\mss$ estimates, which are
quite different. Panel (b) scales these estimates as $\widehat{Z}_y$,
and shows that most joint $\widehat{Z}_y$ estimates are greater than pointwise $\widehat{Z}_y$
estimates (shown in {\color{blue}blue}; exceptions in {\color{red}red}), consistent with the relatively
high posterior estimate for $\rho$ (mean 0.86, 95\% CI 0.82-0.89).}
\label{fig:pairwise-cv-imm}
\end{figure}

Childhood vaccinations are a crucial public health intervention for
preventing the spread of preventable diseases. However, between 2002-2013
Australian registered vaccination objection rates increased from 1.1\%
to 2.0\% \parencite{beardTrendsPatternsVaccination2016}, consistent
with an international pattern of rising vaccine hesitancy \parencite{chanPowerVaccinesStill2017}.
Qualitative research suggests families' objections were clustered
in regional (non-urban) areas, and that socioeconomic status and barriers
to accessing the health system are both factors \parencite{beardTrendsPatternsVaccination2016}.
These stylized facts suggest that vaccination objection is a community-level
phenomenon, and that a community-level analysis is useful for directing
health interventions.

In response to families that fail to vaccinate their children, which
tend to cluster in certain geographic areas (Figure~C.2 in the Appendix) and are not usually related
to religious objections, some states have taken steps to exclude unvaccinated
children from childcare centers and kindergartens, and to lose access to welfare
services \parencite{kirbyNoJabNo2017}. These policy measures, so-called
`no jab no play' laws, have increased overall vaccination rates
\parencite{liRemovingConscientiousObjection2021}. Nonetheless, despite
recent progress, vaccination coverage rates for 5-year-old children
stood at 94.04 per cent as of September 2023, short of the national
target of 95 per cent, with vaccination rates for communities with
lower socio-economic status reporting lower vaccination rates \parencite{doh2023}. 

In this example, we construct an ecological model of non-vaccination
rates for 5-year-old children in Australian communities, in the spirit
of \citeauthor{MAREK2020113292}'s (\citeyear{MAREK2020113292}) study
of children in New Zealand. Healthy Australian children should have
received all scheduled early childhood vaccinations by this age. We
obtained the number of registered children and children whose vaccination
status is not up-to-date in 2021 for 1156 Australian population health
areas (PHAs) from the Social Health Atlas of Australia \parencite{publichealthinformationdevelopmentunitSocialHealthAtlas2024}.
The PHAs are 1165 relatively small geographic areas with a median
estimated population of 19,913 persons. Viewing the decision to comply
with the vaccination schedule as a binary choice, we model the number
of unvaccinated children as (conditionally independent) binomial variables,
where the number of binomial trials $\left(n_{i}\right)$ is the number
of children aged 5 in that PHA and $\left(y_{i}\right)$ the number
of those children that are fully-vaccinated.

In addition to geographical location, we also include three sets of
explanatory variables that are also available at the PHA level. These
are not directly related to vaccine uptake but are instead interpreted
as proxies for health behaviors, socio-economic disadvantage, and
labor market participation (see Appendix~C.1 for
a complete listing).

% The models under consideration are therefore binomial models for each spatial
% region $i$, with linear predictor
% \begin{equation}
% \eta_i = \beta^\top x_i + z_i,\quad i=1, \dots, 1156,
% \end{equation}
% where $z_i$ is a Gaussian random effect, $\beta$ is a vector of fixed effects,
% and $x_i$ is a vector of covariates specific to region $i$. The choice of the 
% particular variables to include in $x_i$ and the joint covariance structure for
% $\left(z_i\right)$, including the choice of network structure, are specific to the 
% candidate model.

To conduct model selection, we use CV to compare a total of 15 spatial
models using spatial cross-validation, which combine various spatial
weight matrixes, covariates, and model formulations (see Appendix~C.1 for a detailed list). This yields
a total of $\begin{psmallmatrix}15\\2\end{psmallmatrix}=105$ pairwise
model comparisons. Non-spatial models are excluded from
the comparison because of the evident spatial distribution of the
data (see e.g. \cite{kuhnLessEightHalf2012}). We employ 12-fold spatially-clustered
CV with a 1-step buffer, computed using the \texttt{spatialsample}
R package (\cite{mahoneyAssessingPerformanceSpatial2023}; Figure~C.1 in the appendix). 
Alternative fold counts and choices for the buffer size
do not significantly change our results.

Joint and pointwise CV selected the same candidate (Model~1, see Table~C.1
in the appendix). The preferred model
is a spatial lag model that includes all proposed explanatory variables,
which suggests that all three proposed explanations can explain childhood
nonvaccination.

Since the preferred model is a SAR, we can interpret $\rho$ in a
similar manner to the simulations in Section~\ref{sec:Simulations}.
The standardized $\rho$ has a posterior mean of 0.86 (95\% CI 0.82-0.89;
Table~C.2 in the Appendix). The results of Section~3 suggest
that under this relatively high value for $\rho$, which denotes strong
spatial autocorrelation (Figure~\ref{fig:imm-fitted}), jointly-evaluated
CV will develop greater statistical power than CV evaluated pointwise.
Indeed, Figure~\ref{fig:pairwise-cv-imm} shows that relative variability
(measured by $Z$ ratios) for the joint is considerably lower for
all but a handful of pairwise model comparisons.

\subsection{Lung cancer in Pennsylvania\label{subsec:Lung-cancer}}

In this subsection, we present an application to standardized incidence
rates (SIRs) for lung cancer in 67 Pennsylvania counties in 2002.
These data and similar models were presented by Moraga (\citeyear{moraga2019geospatial,moragaSmallAreaDisease2018}),
for which data are available in the \texttt{SpatialEpi} R package
\parencite{kimSpatialEpiMethodsData2023}. In this study, the SIR
is standardized across a total of 16 strata (2 race groups, 2 genders,
and 4 age groups).

Lung cancer is the leading cause of death from cancer in the United
States. It is caused by exposure to tobacco smoke as well as other
environmental causes, such as the carcinogen radon \parencite{ALBERG200729S}.
These causes suggest that both location and population smoking rates
could be informative in explaining lung cancer incidence. For this
reason, we include candidate models that include smoking rates for
each county as an explanatory variable. In addition to SAR models,
we also include the Besag-York-Molli\'e \parencite[BYM;][]{besagBayesianImageRestoration1991}
and BYM2 \parencite{rieblerIntuitiveBayesianSpatial2016} models proposed
by \textcite{moraga2019geospatial}. However, even with these candidates
included, we find CV prefers the SAR model. See Appendix~C.2 for details.

Both pointwise and joint CV agree on the choice of preferred model,
which is a SAR that includes only an intercept (Table~C.5 in the Appendix).
The preferred model does not include the smoking variable. (This does
not imply that smoking is not related to lung cancer incidence at
the individual level; rather that variability in smoking rates across
small areas does not explain the epidemiology of smoking incidence;
an example of the ecological fallacy; \cite{cressie2015statistics}.)

\begin{figure}
\includegraphics[width=1\textwidth]{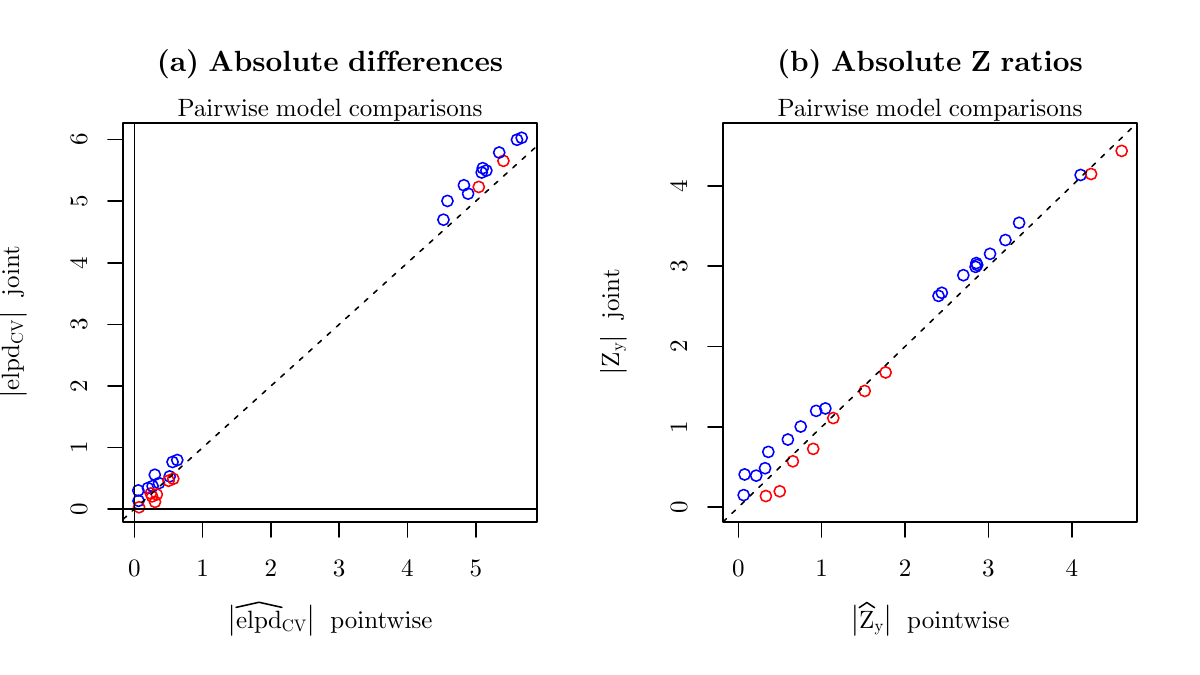}

\caption{Summary of pairwise comparisons among 6 candidate spatial models for
the Pennsylvania lung cancer example (Section~\ref{subsec:Lung-cancer}).
Panel (a) compares the magnitude of pointwise and joint $\mss$ estimates, which are
very similar. Panel (b) scales these estimates as $\widehat{Z}_y$
and shows $\widehat{Z}_y$ estimates clustered around the $45^\circ$ line; roughly
half of estimates are above and below the line, shown respectively in {\color{blue}blue} and {\color{red}red},
consistent with the relatively low estimate for $\rho$ found in Table~\ref{penn-coef-est}.}
\label{penn-coef-est}
\end{figure}

Under the preferred model, $\rho$ has a mean posterior estimate of
0.58 (CI 0.45 - 0.71; Table~C.5 in the Appendix). This is
a relatively low value, and we would expect there to be little difference
between pointwise and joint CV estimates. The similarity between these
two quantities is evident in Figure~\ref{penn-coef-est}, which compares
pairwise model comparisons for jointly- and pointwise-computed CV
estimates.

\section{Discussion and conclusion\label{sec:Conclusion}}

We have extended earlier work on cross-validatory
model selection for models of dependent data to spatial dependence
structures and approximate inference methods (Laplace approximation). In this paper our analysis goes beyond
selecting just the regression part of the model for univariate autoregressions,
and we have illustrated cases where jointly-evaluated CV is useful for selecting
other aspects of the model.

In our experiments, when spatial dependence is strong, spatial
CV approaches with relatively small numbers of folds (a dozen or so)
but larger test set sizes develop greater model selection power 
for selecting between spatial models than
those with smaller test sets, such as LOO-CV.
We have found the $Z$ ratio as a useful lens for summarizing the
difference between evaluation methods, since pointwise and joint 
selection statistic distributions differ in both location and variance.

% The underlying reasons why joint evaluation methods perform better under
% strong dependence is not yet fully clear. We speculate that
% when dependence is strong, uncertainty about the covariance structure is
% greater.

Our experiments are subject to several limitations. First, we
consider only comparisons between \emph{spatial} models. We have not
considered comparisons invovling \emph{iid} models, where spatial covariances
are not explicitly modeled. Second, we have considered only the logarithmic
scoring rule. While this is by far the most popular multivariate scoring rule for
probabilistic models because of its deep connection to statistical
concepts of entropy and Kullback-Liebler divergence \parencite{dawidBayesianModelSelection2015},
a range of alternatives are available \parencite{gneitingStrictlyProperScoring2007}.

Perhaps the most significant limitation is that we have interpreted our
results in terms of the standardized $\rho$ parameter in a SAR with row-standardized
weights, which is bounded on $[0,1)$ (for models with nonnegative autocorrelation,
which is the usual case.)
However, different spatial models account for spatial dependence
in ways that are not comparable to standardized $\rho$: for example,
when $W$ is not row-standardized, $\rho$ is bounded by $\left(1/\omega_{-},1/\omega_{+}\right)$,
for $\omega_{-}$ and $\omega_{+}$ respectively the smallest and
largest eigenvalues of the adjacency matrix. So-called `intrinsic'
models \parencite{besagBayesianImageRestoration1991,rueGaussianMarkovRandom2005,cressie2015statistics}
eliminate such parameters altogether. Further investigation is needed
to determine a model-independent measure of dependence: one potential
approach might be to measure $\rho$ for the problem using an encompassing
SAR model, regardless of the model preferred by CV.

\subsection*{Acknowledgments}

AC's work was supported in part by an Australian Government Research
Training Program Scholarship. AV acknowledges the Research Council
of Finland Flagship program: Finnish Center for Artificial Intelligence,
and Academy of Finland project (340721). CF acknowledges financial
support under National Science Foundation Grant SES-1921523.

\printbibliography[heading=bibintoc]

\newpage
\appendix
\numberwithin{equation}{section}
\numberwithin{figure}{section}
\numberwithin{table}{section}

\section{Supplementary experiments}

The simulations presented in this section complement experiments in
Section~3 in the main text.

% \subsection{Larger sample for covariate selection}

% This experiment extends Subsection~\ref{subsec:SAR-p} by performing
% model selections on a much larger lattice. For simplicity, we present
% only $\Delta$ estimates produced using joint evaluation.

% \begin{figure}[H]
% \begin{centering}
% \includegraphics[width=1\textwidth]{figures/ex3_2panel_by_side}\\
% \includegraphics[width=1\textwidth]{figures/ex3_intervals_by_rhoxside_2}
% \par\end{centering}
% \caption{SAR model selection experiment on a 60x60 grid. Rook contiguity.}
% \label{fig:sar-beta-sel-large}
% \end{figure}

\subsection{SAR network structure selection (queen dgp)\label{subsec:SAR-W-queen}}

\begin{figure}[H]
\begin{centering}
\includegraphics[width=1\textwidth]{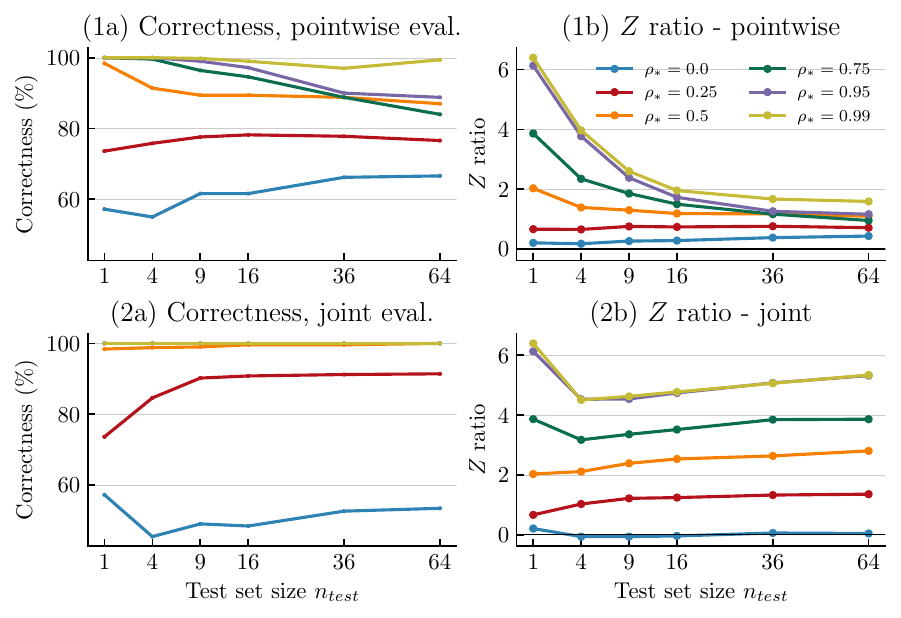}
\par\end{centering}
\caption{Summary of a sequence of model selection experiments to 
identify the model network, each with 500 independent replications, for 
SAR models a regular 
lattice with $n=576$. Model selection is by blockwise CV with the block
size noted on the $x$-axis. The model and dgps are Gaussian SARs, with
$\left(I_n - \rho W_+\right)y=X\beta + \sigma\varepsilon$.
True $\sigma_*^2=4$, $\beta_*=(1,1,1)$ and $X$ 
is a $n\times3$ matrix containing a column of ones and 
standard normal draws. Candidate models are distinguished by $W_+$,
the row-standardized agency matrix.
The dgp has queen adjacency and candidate models include rook
and queen adjacency (see Figure~1 in the main text). Covariates are
common to dgp and both candidates. Colors indicate the true $\rho_*$
parameter. Panels (1a) and (2a) plot the share
of independent data draws where the correct model is selected. Panels
(1b) and (2b) plot the $Z$-ratio of the resulting distribution
of model selection statistics $\mss$, oriented so that positive values
denote correct model selection. Panels (1a) and (1b) are computed using
pointwise evaluation; (2a) and (2b) joint evaluation. Except the case where $\rho_*=0$, joint evaluation yields higher
accuracy and lower relative variability ($Z$ ratio). See also Figure~\ref{lbl:sar-connectivity-sel-dist-queen}.}
\label{lbl:sar-connectivity-sel-1}
\end{figure}

\begin{figure}[H]
\begin{centering}
\includegraphics[width=1\textwidth]{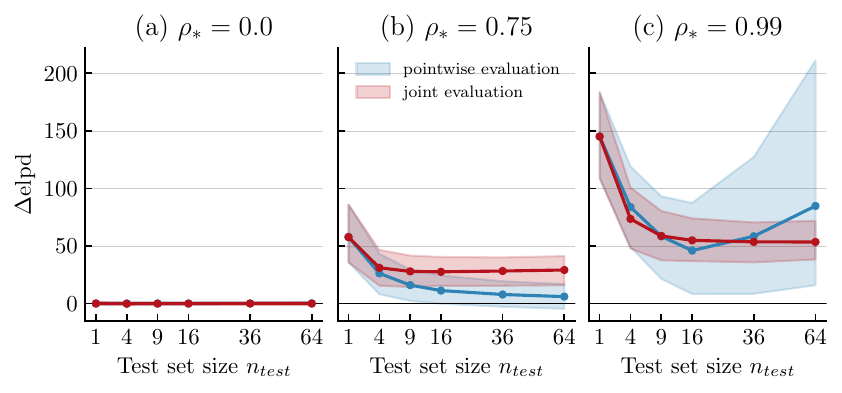}
\par\end{centering}
\caption{Comparison of 90\% intervals for the distribution of $\mss$ over 500 independent $y$ draws, for sequences of SAR network structure selection experiments on a regular lattice with $n=576$ (Appendix~A.1). Each panel shows intervals representing pointwise (blue) and joint (red) model selection statistics.
Probability mass above the $x$-axis represents correct model selection, and vice versa.
Panels (a)-(c) respectively show an increasing degree of dependence $\rho_*$ in the true
dgp. For $\rho_*=0$, the variance for both measures is small and covers the $x$-axis, indicating poor selection performance for both. For $\rho_*>0$, selection performance 
increases for multivariate test sets, indicated by greater probability mass in the first quadrant.
However, notice the sharp relative increase in variance for the pointwise measure as 
$\rho_*$ increases, and for $\rho_*>0$, as the test set size increases, along with a 
downward location shift toward the $x$-axis for the pointwise measure.}
\label{lbl:sar-connectivity-sel-dist-queen}
\end{figure}

\subsection{Model network degree selection\label{subsec:SAR-p}}

\begin{figure}[H]
\begin{centering}
\includegraphics[width=1\textwidth]{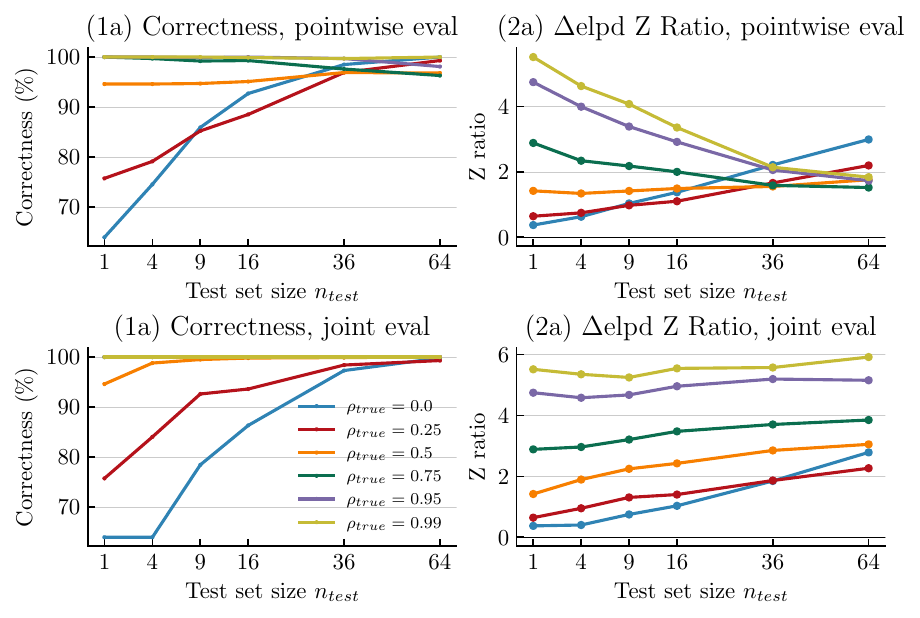}
\par\end{centering}
\caption{Summary of a sequence of model selection experiments to 
identify the appropriate contiguity degree (number of steps treated as neighbors) for a SAR model. In these experiments, the dgp is a SAR(1), which means that neighbors are a single
step away in the contiguity network. The alternative model is a SAR(2). Each experiment includes 1,000 independent replications, for SAR models on a regular 
lattice with $n=576$. Model selection is by blockwise CV with the block
size noted on the $x$-axis. The model and dgps are Gaussian SARs, with
$\left(I_n - \rho W_+\right)y=X\beta + \sigma\varepsilon$.
True $\sigma_*=5$ and $X$ 
is a $n\times2$ matrix containing a column of ones and a column of 
standard normal draws. Candidate models are distinguished by $W_+$,
the row-standardized agency matrix.
Covariates are common to dgp and both candidates. Colors indicate the true $\rho_*$
parameter. Panels (1a) and (2a) plot the share
of independent data draws where the correct model is selected. Panels
(1b) and (2b) plot the $Z$-ratio of the resulting distribution
of model selection statistics $\mss$, oriented so that positive values
denote correct model selection. Panels (1a) and (1b) are computed using
pointwise evaluation; (2a) and (2b) joint evaluation. Except for the case where $\rho_*=0$, joint evaluation yields higher
accuracy and lower relative variability ($Z$ ratio). See also Figure~\ref{lbl:sar-degree-sel-dist}.}
\label{lbl:sar-degree-sel-1}
\end{figure}

\begin{figure}[H]
\begin{centering}
\includegraphics[width=1\textwidth]{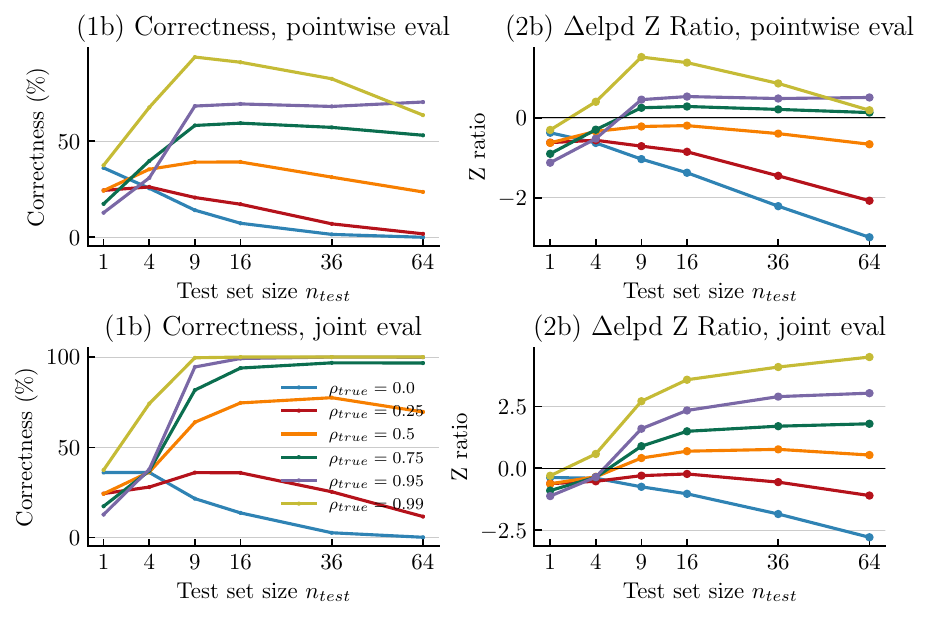}
\par\end{centering}
\caption{Summary of a sequence of model selection experiments to 
identify the appropriate contiguity degree (number of steps treated as neighbors) for a SAR model. In these experiments, the dgp is a SAR(2), i.e. neighbors are two steps away in the lattice. The alternative model is a SAR(1). Each experiment includes 1,000 independent replications, for SAR models a regular 
lattice with $n=576$. Model selection is by blockwise CV with the block
size noted on the $x$-axis. The model and dgps are Gaussian SARs, with
$\left(I_n - \rho W_+\right)y=X\beta + \sigma\varepsilon$.
True $\sigma_*=5$ and $X$ 
is a $n\times2$ matrix containing a column of ones and a column of 
standard normal draws. Candidate models are distinguished by $W_+$,
the row-standardized agency matrix.
Covariates are
common to dgp and both candidates. Colors indicate the true $\rho_*$
parameter. Panels (1a) and (2a) plot the share
of independent data draws where the correct model is selected. Panels
(1b) and (2b) plot the $Z$-ratio of the resulting distribution
of model selection statistics $\mss$, oriented so that positive values
denote correct model selection. Panels (1a) and (1b) are computed using
pointwise evaluation; (2a) and (2b) joint evaluation. Except for the case where $\rho_*=0$, joint evaluation yields higher
accuracy and lower relative variability ($Z$ ratio). See also Figure~\ref{lbl:sar-degree-sel-dist}.}
\label{lbl:sar-degree-sel-2}
\end{figure}

\begin{figure}[H]
\begin{centering}
\includegraphics[width=1\textwidth]{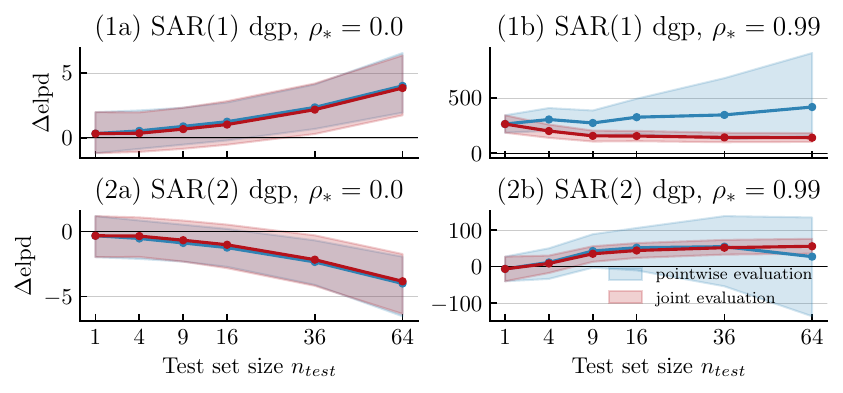}
\par\end{centering}
\caption{Comparison of 90\% intervals for the distribution of $\mss$ over 1,000 independent $y$ draws, for sequences of SAR order selection experiments on a regular lattice with $n=576$. Each panel shows intervals representing pointwise (blue) and joint (red) model selection statistics.
Probability mass above the $x$-axis represents correct model selection, and vice versa. Note that our focus is the relative performance of pointwise
and blue model selection statistics, not absolute performance.
When $\rho_*=0$, there is little difference between joint and pointwise 
methods, either in location or variance.
For strong dependence ($\rho_*=0.99$; Panels (1b) and (2b)), the variance of 
pointwise methods increases sharply, leading to worse model selection 
performance (Figures~\ref{lbl:sar-degree-sel-1} and
\ref{lbl:sar-degree-sel-2}). In this figure, this is visible in Panel (2b) as 
probability mass appearing in the fourth quadrant.}
\label{lbl:sar-degree-sel-dist}
\end{figure}

\newpage
\section{Additional figures}

These figures complement experiments presented in Section~3 of the main text.

\begin{figure}[H]
\begin{centering}
\includegraphics[width=1\textwidth]{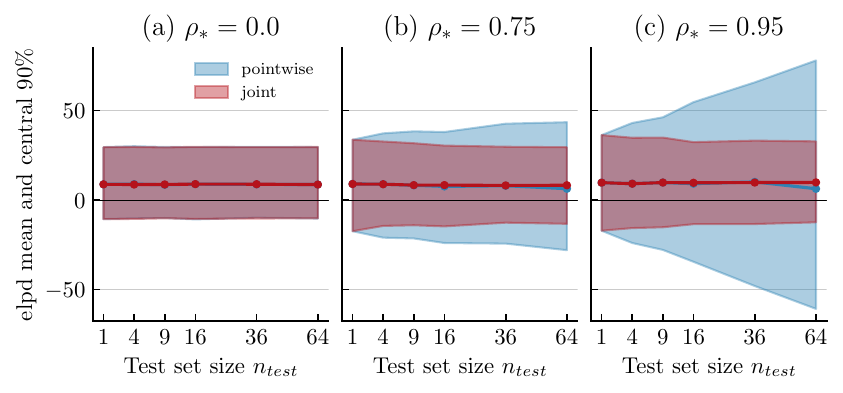}
\par\end{centering}
\caption{Comparison of 90\% intervals for $\mss$ for sequences of SAR covariate 
selection experiments on a regular lattice with $n=576$ (Section~3.1.1, main text). Each 
panel shows intervals representing pointwise (blue) and joint (red) model selection 
statistics.
Probability mass above the $x$-axis represents correct model selection, and vice versa.
Panels (a)-(c) respectively show an increasing degree of dependence $\rho_*$ in the true
dgp. Notice the relative increase in variance for the pointwise measure as $\rho_*$
increases, and for $\rho_*>0$, as the test set size increases, along with a moderate
location shift toward the $x$-axis for the pointwise measure.}
\label{fig:sar-beta-dist}
\end{figure}

\begin{figure}[H]
\begin{centering}
\includegraphics[width=1\textwidth]{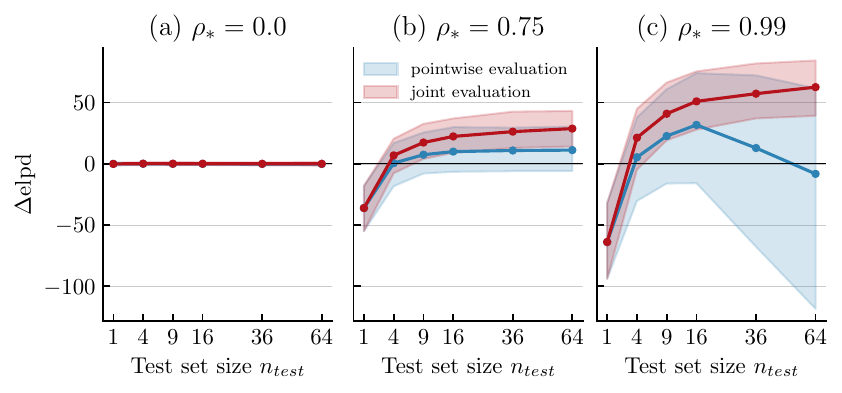}
\par\end{centering}
\caption{Comparison of 90\% intervals for $\mss$ for sequences of SAR network structure selection experiments on a regular lattice with $n=576$ (Section~3.1.2, main text). Each panel shows intervals representing pointwise (blue) and joint (red) model selection statistics.
Probability mass above the $x$-axis represents correct model selection, and vice versa.
Panels (a)-(c) respectively show an increasing degree of dependence $\rho_*$ in the true
dgp. For $\rho_*=0$, the variance for both measures is small and covers the $x$-axis, indicating poor selection performance for both. For $\rho_*>0$, selection performance 
increases for multivariate test sets, indicated by greater probability mass in the first quadrant.
However, notice the sharp relative increase in variance for the pointwise measure as 
$\rho_*$ increases, and for $\rho_*>0$, as the test set size increases, along with a 
downward location shift toward the $x$-axis for the pointwise measure.}
\label{lbl:sar-connectivity-sel-dist-rook}
\end{figure}

\begin{figure}[H]
\begin{centering}
\includegraphics[width=1\textwidth]{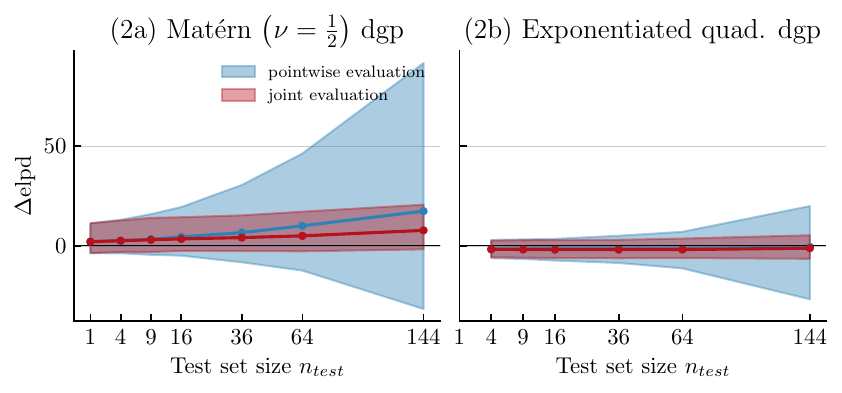}
\par\end{centering}
\caption{Comparison of 90\% intervals for the distribution of $\mss$ over 500 
independent $y$ draws, for sequences of kernel selection experiments on a 
regular lattice with $n=576$ (Section~3.2, main text). Each panel shows 
intervals representing pointwise (blue) and joint (red) model selection 
statistics. Correct model selection is indicated by probability mass in the
first quadrant, incorrect selection in the fourth quadrant.}

\label{lbl:materneq:correct-dist}
\end{figure}

\section{Details of case studies}

\subsection{Australian child vaccination model\label{sec:vax-supp}}

\begin{figure}[H]
\begin{centering}
\includegraphics[width=1\textwidth]{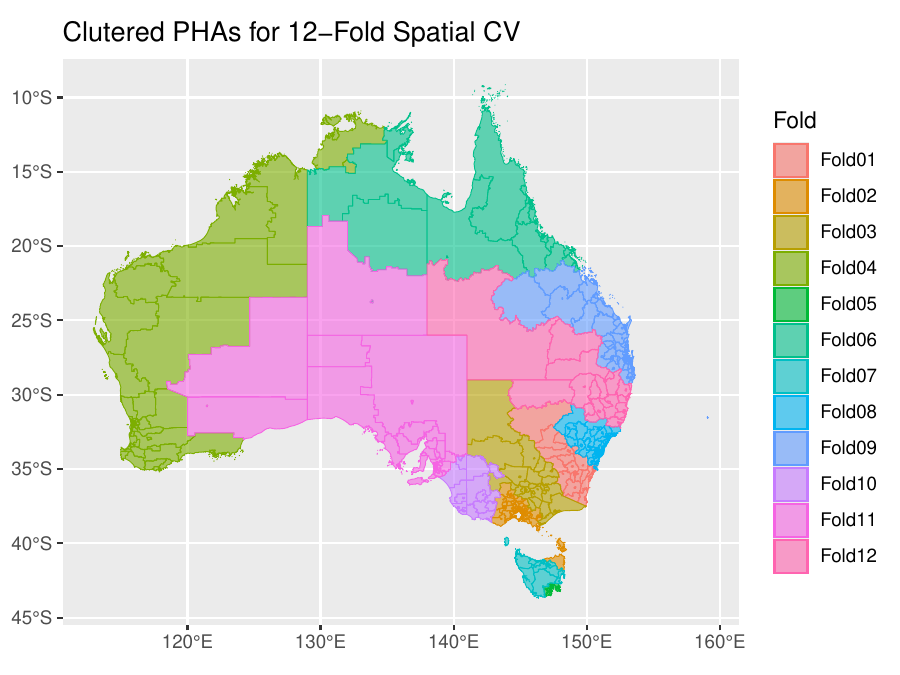}
\par\end{centering}
\caption{12 spatially-clustered CV folds for population health areas (PHAs; \cite{publichealthinformationdevelopmentunitPopulationHealthAreas2023}) computed using the spatialsample package
\parencite{mahoneyAssessingPerformanceSpatial2023}.}
\label{fig:imm-clust}
\end{figure}

\subsubsection{Data}

For 1156 PHAs, we obtain the number of children fully vaccinated at age five and the total number
of registered children aged five in 2021 using data from the Social Health Atlas of Australia.
The number of unvaccinated children is calculated as the difference between these numbers.

The following variables are sourced at the PHA level from the Australian
Social Atlas \parencite{publichealthinformationdevelopmentunitSocialHealthAtlas2024}.
\begin{description}
%\item [{remoteness}] 2021 Australian Statistical Geography Standard (ASGS)
%remoteness areas \parencite{australianbureauofstatisticsAustralianStatisticalGeography2021},
%computed for 2021 PHAs \parencite{publichealthinformationdevelopmentunitPopulationHealthAreas2023}
%with QGIS software version 3.38 \parencite{qgis.orgQGISGeographicInformation2024}.
\item [{fully\_breastfed\_6m\_pc}] Share of children fully breastfed at
six months of age
\item [{pc\_nbcsp\_part}] Share of eligible individuals electing to participate
in the National Bowel Cancer Screening Program
\item [{pc\_est\_daily\_drink}] Estimated share of individuals who drink
alcohol daily
\item [{unpaid\_childcare}] Share of individuals over
age 15 whose main occupation is unpaid childcare for their own children
\item [{preschool\_5yo\_pc}] Share of registered 5-year-old children enrolled
at a preschool
\item [{pc\_early\_school\_leaver}] Share of adults that left school before
year 12
\item [{pc\_ft\_school\_age\_16}] Share of children aged 16 enrolled in
school full-time
\item [{pc\_unemp}] Estimated unemployment rate
\item [{pc\_part\_rate}] Estimated labor force participation rate
\item [{pc\_private\_health\_ins}] Estimated share of individuals with
private health insurance
\item [{seifa\_disadv\_index}] Index of Relative Socio-Economic Disadvantage,
from the Socio-Economic Indexes for Areas (SEIFA) \parencite{australianbureauofstatisticsSocioEconomicIndexesAreas2021}
\item [{pc\_financial\_stress\_rent}] Estimated share of households in
financial distress due to rent
\item [{pc\_financial\_stress\_mtg}] Estimated share of households in financial
distress due to mortgage payments
\item [{low\_inc\_hholds}] Estimated share of low-income households
\item [{pc\_crowded\_dwellings}] Estimated share of overcrowded houses
\item [{pc\_child\_jobless\_family}] Estimated share of households with
children where no parent is employed
\item [{pc\_moth\_lowed}] Estimated share of children whose
mother has less than a year 12 education attainment
\end{description}

\subsubsection{Candidate models and results\label{subsec:imm-cand}}

For all models, the observation density, indexed by PHA $i$, is
$$p\left(y_{i}\,|\,p_{i}\right) = \mathcal{B}\left(y_{i}\,|\,n_{i},S\left(\lambda_{i}\right)\right), \label{eq:vax-obs}$$
where $\mathcal{B}\left(y\,|\,n,p\right)$ is the binomial density, $S\left(\cdot\right)$ is the sigmoid link
function.

Table~\ref{tbl:imm-cov} summarizes the candidate models.
The candidate models differ in three respects: the adjacency weights $W$, the functional
form of the model for the latent states $z$, and the covariates $X$.

The latent quantity $z$ is a GMRF with sparsity structure defined by $W$
$$p\left(z\,|\,\theta\right) =\mathcal{N}\left(x_{i}^{\top}\beta\,|\,0,\Sigma_{\theta}\right).\label{eq:vax-spatial}
$$
We propose the standard SAR model \parencite{anselin1988spatial}:
\begin{equation}
z=\left(I_{n}-\rho W\right)^{-1}\left(X\beta+\tau^{-1}\varepsilon\right)\label{eq:imm-cand-1}
\end{equation}
In addition, we also include the following modified SAR \parencite{kisslingSpatialAutocorrelationSelection2008}:
\begin{equation}
z=X\beta + \left(I_{n}-\rho W\right)^{-1}\tau^{-1}\varepsilon\label{eq:imm-cand-2}
\end{equation}

To close the model, we impose weakly informative
priors $\beta\sim\mathcal{N}\left(0,10\cdot I_{k}\right)$
and $\rho\sim \mathcal{B}\left(2,2\right)$.

\begin{table}[H]
\begin{centering}
\begin{turn}{90}
\begin{tabular}{llccccccccccccccc}
\hline 
\multirow{2}{*}{} &  & \multicolumn{15}{c}{\textbf{Candidate model}}\tabularnewline
 &  & $M_1$ & $M_2$ & $M_3$ & $M_4$ & $M_5$ & $M_6$ & $M_7$ & $M_8$ & $M_9$ & $M_{10}$ & $M_{11}$ & $M_{12}$ & $M_{13}$ & $M_{14}$ & $M_{15}$\tabularnewline
\cline{1-1} \cline{3-17} \cline{4-17} \cline{5-17} \cline{6-17} \cline{7-17} \cline{8-17} \cline{9-17} \cline{10-17} \cline{11-17} \cline{12-17} \cline{13-17} \cline{14-17} \cline{15-17} \cline{16-17} \cline{17-17} 
\textbf{Specification} &  & \ref{eq:imm-cand-1} & \ref{eq:imm-cand-1} & \ref{eq:imm-cand-1} & \ref{eq:imm-cand-1} & \ref{eq:imm-cand-1} & \ref{eq:imm-cand-2} & \ref{eq:imm-cand-2} & \ref{eq:imm-cand-2} & \ref{eq:imm-cand-2} & \ref{eq:imm-cand-2}  & \ref{eq:imm-cand-1} & \ref{eq:imm-cand-1} & \ref{eq:imm-cand-1} & \ref{eq:imm-cand-1} & \ref{eq:imm-cand-1} \tabularnewline
\textbf{Weights} &  & $W_{+}$ & $W_{+}$ & $W_{+}$ & $W_{+}$ & $W_{+}$ & $W_{+}$ & $W_{+}$ & $W_{+}$ & $W_{+}$ & $W_{+}$ & $W$ & $W$ & $W$ & $W$ & $W$ \tabularnewline
\textbf{Covariates} &  &  &  &  &  &  &  &  &  &  &  &  &  &  &  & \tabularnewline
fully\_breastfed\_6m\_pc &  & \Checkmark{} &  & \Checkmark{} &  &  & \Checkmark{} &  & \Checkmark{} &  &  & \Checkmark{} &  & \Checkmark{} &  & \tabularnewline
pc\_nbcsp\_part &  & \Checkmark{} &  & \Checkmark{} &  &  & \Checkmark{} &  & \Checkmark{} &  &  & \Checkmark{} &  & \Checkmark{} &  & \tabularnewline
pc\_est\_daily\_drink &  & \Checkmark{} &  & \Checkmark{} &  &  & \Checkmark{} &  & \Checkmark{} &  &  & \Checkmark{} &  & \Checkmark{} &  & \tabularnewline
unpaid\_childcare &  & \Checkmark{} & \Checkmark{} &  &  & \Checkmark{} & \Checkmark{} & \Checkmark{} &  &  & \Checkmark{} & \Checkmark{} & \Checkmark{} &  &  & \Checkmark{}\tabularnewline
preschool\_5yo\_pc &  & \Checkmark{} &  &  &  &  & \Checkmark{} &  &  &  &  & \Checkmark{} &  &  &  & \tabularnewline
pc\_early\_school\_leaver &  & \Checkmark{} &  &  &  &  & \Checkmark{} &  &  &  &  & \Checkmark{} &  &  &  & \tabularnewline
pc\_ft\_school\_age\_16 &  & \Checkmark{} & \Checkmark{} &  & \Checkmark{} &  & \Checkmark{} & \Checkmark{} &  & \Checkmark{} &  & \Checkmark{} & \Checkmark{} &  & \Checkmark{} & \tabularnewline
pc\_unemp &  & \Checkmark{} & \Checkmark{} &  &  & \Checkmark{} & \Checkmark{} & \Checkmark{} &  &  & \Checkmark{} & \Checkmark{} & \Checkmark{} &  &  & \Checkmark{}\tabularnewline
pc\_part\_rate &  & \Checkmark{} & \Checkmark{} &  &  & \Checkmark{} & \Checkmark{} & \Checkmark{} &  &  & \Checkmark{} & \Checkmark{} & \Checkmark{} &  &  & \Checkmark{}\tabularnewline
pc\_private\_health\_ins &  & \Checkmark{} & \Checkmark{} & \Checkmark{} &  &  & \Checkmark{} & \Checkmark{} & \Checkmark{} &  &  & \Checkmark{} & \Checkmark{} & \Checkmark{} &  & \tabularnewline
seifa\_disadv\_index &  & \Checkmark{} &  &  & \Checkmark{} &  & \Checkmark{} &  &  & \Checkmark{} &  & \Checkmark{} &  &  & \Checkmark{} & \tabularnewline
pc\_financial\_stress\_rent &  & \Checkmark{} & \Checkmark{} &  & \Checkmark{} &  & \Checkmark{} & \Checkmark{} &  & \Checkmark{} &  & \Checkmark{} & \Checkmark{} &  & \Checkmark{} & \tabularnewline
pc\_financial\_stress\_mtg &  & \Checkmark{} & \Checkmark{} &  & \Checkmark{} &  & \Checkmark{} & \Checkmark{} &  & \Checkmark{} &  & \Checkmark{} & \Checkmark{} &  & \Checkmark{} & \tabularnewline
low\_inc\_hholds &  & \Checkmark{} & \Checkmark{} &  & \Checkmark{} &  & \Checkmark{} & \Checkmark{} &  & \Checkmark{} &  & \Checkmark{} & \Checkmark{} &  & \Checkmark{} & \tabularnewline
pc\_crowded\_dwellings &  & \Checkmark{} & \Checkmark{} &  & \Checkmark{} &  & \Checkmark{} & \Checkmark{} &  & \Checkmark{} &  & \Checkmark{} & \Checkmark{} &  & \Checkmark{} & \tabularnewline
pc\_child\_jobless\_family &  & \Checkmark{} & \Checkmark{} &  &  & \Checkmark{} & \Checkmark{} & \Checkmark{} &  &  & \Checkmark{} & \Checkmark{} & \Checkmark{} &  &  & \Checkmark{}\tabularnewline
pc\_moth\_lowed &  & \Checkmark{} &  &  &  &  & \Checkmark{} &  &  &  &  & \Checkmark{} &  &  &  & \tabularnewline
\hline 
\end{tabular}
\end{turn}
\par\end{centering}
\caption{Candidate model specifications. Weights matrixes $W$ and $W_+$ are defined in equations (8) and (9) in the main text.}
\label{tbl:imm-cov}
\end{table}

\begin{table}
\begin{centering}
\begin{tabular}{llr@{\extracolsep{0pt}.}lr@{\extracolsep{0pt}.}lr@{\extracolsep{0pt}.}lr@{\extracolsep{0pt}.}lr@{\extracolsep{0pt}.}l}
\hline 
Parameter &  & \multicolumn{2}{c}{mean} & \multicolumn{2}{c}{sd} & \multicolumn{2}{c}{$q_{0.025}$} & \multicolumn{2}{c}{$q_{0.5}$} & \multicolumn{2}{c}{$q_{0.975}$}\tabularnewline
\cline{1-1} \cline{3-12} \cline{5-12} \cline{7-12} \cline{9-12} \cline{11-12} 
Precision ($\tau$) &  & 17&0 & 2&1 & 13&3 & 16&9 & 21&6\tabularnewline
Spatial parameter ($\rho$) &  & 0&86 & 0&017 & 0&82 & 0&86 & 0&89\tabularnewline
Regression coef. ($\beta$) &  & \multicolumn{2}{c}{} & \multicolumn{2}{c}{} & \multicolumn{2}{c}{} & \multicolumn{2}{c}{} & \multicolumn{2}{c}{}\tabularnewline
\quad{}(Intercept) &  & -6&32 & 10&31 & -26&64 & -6&32 & 13&98\tabularnewline
\noalign{\vskip6pt}
\multicolumn{12}{l}{\quad{}Health behavior}\tabularnewline
\quad{}\quad{}fully\_breastfed\_6m\_pc &  & 0&003 & 0&002 & -0&001 & 0&003 & 0&008\tabularnewline
\quad{}\quad{}pc\_nbcsp\_part &  & -0&001 & 0&000 & -0&002 & -0&001 & 0&000\tabularnewline
\quad{}\quad{}pc\_est\_daily\_drink &  & 0&021 & 0&005 & 0&011 & 0&021 & 0&031\tabularnewline
\quad{}\quad{}pc\_private\_health\_ins &  & -0&006 & 0&002 & -0&010 & -0&006 & -0&001\tabularnewline
\noalign{\vskip6pt}
\multicolumn{12}{l}{\quad{}Socioeconomic disadvantage}\tabularnewline
\quad{}\quad{}seifa\_disadv\_index &  & 0&004 & 0&001 & 0&002 & 0&004 & 0&007\tabularnewline
\quad{}\quad{}pc\_financial\_stress\_rent &  & 0&009 & 0&004 & 0&001 & 0&009 & 0&016\tabularnewline
\quad{}\quad{}pc\_financial\_stress\_mtg &  & 0&044 & 0&008 & 0&028 & 0&044 & 0&060\tabularnewline
\quad{}\quad{}pc\_early\_school\_leaver &  & -0&014 & 0&005 & -0&024 & -0&014 & -0&004\tabularnewline
\quad{}\quad{}low\_inc\_hholds &  & -0&009 & 0&005 & -0&018 & -0&009 & 0&000\tabularnewline
\quad{}\quad{}pc\_crowded\_dwellings &  & 0&034 & 0&006 & 0&022 & 0&034 & 0&046\tabularnewline
\quad{}\quad{}pc\_moth\_lowed &  & 0&006 & 0&007 & -0&007 & 0&006 & 0&018\tabularnewline
\noalign{\vskip6pt}
\multicolumn{12}{l}{\quad{}Education and labor market participation}\tabularnewline
\quad{}\quad{}pc\_child\_jobless\_family &  & -0&006 & 0&007 & -0&020 & -0&006 & 0&008\tabularnewline
\quad{}\quad{}pc\_part\_rate &  & -0&005 & 0&003 & -0&011 & -0&005 & 0&001\tabularnewline
\quad{}\quad{}pc\_unemp &  & 0&034 & 0&012 & 0&010 & 0&034 & 0&057\tabularnewline
\quad{}\quad{}pc\_ft\_school\_age\_16 &  & -0&006 & 0&003 & -0&011 & -0&006 & 0&000\tabularnewline
\quad{}\quad{}preschool\_5yo\_pc &  & -0&007 & 0&002 & -0&011 & -0&007 & -0&002\tabularnewline
\quad{}\quad{}unpaid\_childcare &  & -0&032 & 0&005 & -0&041 & -0&032 & -0&023\tabularnewline
\hline 
\end{tabular}
\par\end{centering}
\caption{Summary of posterior marginals for preferred model for the Australian
child non-vaccination rates example (Section~4.1, main text).
Note the relatively large magnitude of the spatial parameter $\rho$, with a posterior mean of 
$0.86$. Our simulation
study suggests that spatial dependence is strong enough for the joint and pointwise 
model selection statistic estimates
(respectively, $\widehat{\mathrm{elpd}}_{CV}^{j}\left(M_{A},M_{B}\,|\,y\right)$ and $\widehat{\mathrm{elpd}}_{CV}^{pw}\left(M_{A},M_{B}\,|\,y\right)$) to be dissimilar.}
\label{imm-prefer-coef}
\end{table}

\begin{table}[H]
\begin{centering}
\begin{turn}{90}
\begin{tabular}{cr@{\extracolsep{0pt}.}lr@{\extracolsep{0pt}.}lr@{\extracolsep{0pt}.}lr@{\extracolsep{0pt}.}lr@{\extracolsep{0pt}.}lr@{\extracolsep{0pt}.}lr@{\extracolsep{0pt}.}lr@{\extracolsep{0pt}.}lr@{\extracolsep{0pt}.}lr@{\extracolsep{0pt}.}lr@{\extracolsep{0pt}.}lr@{\extracolsep{0pt}.}lr@{\extracolsep{0pt}.}lr@{\extracolsep{0pt}.}lr@{\extracolsep{0pt}.}l}
\hline 
 & \multicolumn{2}{c}{{\small{}$M_{1}$}} & \multicolumn{2}{c}{{\small{}$M_{2}$}} & \multicolumn{2}{c}{\textbf{\small{}$M_{3}$}} & \multicolumn{2}{c}{\textbf{\small{}$M_{4}$}} & \multicolumn{2}{c}{\textbf{\small{}$M_{5}$}} & \multicolumn{2}{c}{\textbf{\small{}$M_{6}$}} & \multicolumn{2}{c}{\textbf{\small{}$M_{7}$}} & \multicolumn{2}{c}{\textbf{\small{}$M_{8}$}} & \multicolumn{2}{c}{\textbf{\small{}$M_{9}$}} & \multicolumn{2}{c}{\textbf{\small{}$M_{10}$}} & \multicolumn{2}{c}{\textbf{\small{}$M_{11}$}} & \multicolumn{2}{c}{\textbf{\small{}$M_{12}$}} & \multicolumn{2}{c}{\textbf{\small{}$M_{13}$}} & \multicolumn{2}{c}{\textbf{\small{}$M_{14}$}} & \multicolumn{2}{c}{\textbf{\small{}$M_{15}$}}\tabularnewline
\hline 
{\small{}$M_{1}$} & \multicolumn{2}{c}{{\small{}---}} & \textcolor{green}{\small{}458}&\textcolor{green}{\small{}1} & \textcolor{green}{\small{}482}&\textcolor{green}{\small{}4} & \textcolor{green}{\small{}425}&\textcolor{green}{\small{}8} & \textcolor{green}{\small{}458}&\textcolor{green}{\small{}8} & \textcolor{green}{\small{}74}&\textcolor{green}{\small{}6} & \textcolor{green}{\small{}645}&\textcolor{green}{\small{}1} & \textcolor{green}{\small{}938}&\textcolor{green}{\small{}6} & \textcolor{green}{\small{}870}&\textcolor{green}{\small{}3} & \textcolor{green}{\small{}872}&\textcolor{green}{\small{}3} & \textcolor{green}{\small{}1}&\textcolor{green}{\small{}7} & \textcolor{green}{\small{}109}&\textcolor{green}{\small{}7} & \textcolor{green}{\small{}450}&\textcolor{green}{\small{}3} & \textcolor{green}{\small{}314}&\textcolor{green}{\small{}6} & \textcolor{green}{\small{}373}&\textcolor{green}{\small{}7}\tabularnewline
{\small{}$M_{2}$} & \textcolor{purple}{\small{}-5}&\textcolor{purple}{\small{}4} & \multicolumn{2}{c}{{\small{}---}} & \textcolor{green}{\small{}24}&\textcolor{green}{\small{}3} & \textcolor{green}{\small{}-32}&\textcolor{green}{\small{}3} & \textcolor{green}{\small{}0}&\textcolor{green}{\small{}7} & \textcolor{green}{\small{}-383}&\textcolor{green}{\small{}5} & \textcolor{green}{\small{}187}&\textcolor{green}{\small{}0} & \textcolor{green}{\small{}480}&\textcolor{green}{\small{}5} & \textcolor{green}{\small{}412}&\textcolor{green}{\small{}2} & \textcolor{green}{\small{}414}&\textcolor{green}{\small{}2} & \textcolor{green}{\small{}-456}&\textcolor{green}{\small{}4} & \textcolor{green}{\small{}-348}&\textcolor{green}{\small{}4} & \textcolor{green}{\small{}-7}&\textcolor{green}{\small{}8} & \textcolor{green}{\small{}-143}&\textcolor{green}{\small{}5} & \textcolor{green}{\small{}-84}&\textcolor{green}{\small{}4}\tabularnewline
\textbf{\small{}$M_{3}$} & \textcolor{purple}{\small{}42}&\textcolor{purple}{\small{}8} & \textcolor{purple}{\small{}48}&\textcolor{purple}{\small{}2} & \multicolumn{2}{c}{{\small{}---}} & \textcolor{green}{\small{}-56}&\textcolor{green}{\small{}6} & \textcolor{green}{\small{}-23}&\textcolor{green}{\small{}6} & \textcolor{green}{\small{}-407}&\textcolor{green}{\small{}8} & \textcolor{green}{\small{}162}&\textcolor{green}{\small{}7} & \textcolor{green}{\small{}456}&\textcolor{green}{\small{}2} & \textcolor{green}{\small{}387}&\textcolor{green}{\small{}9} & \textcolor{green}{\small{}389}&\textcolor{green}{\small{}9} & \textcolor{green}{\small{}-480}&\textcolor{green}{\small{}7} & \textcolor{green}{\small{}-372}&\textcolor{green}{\small{}7} & \textcolor{green}{\small{}-32}&\textcolor{green}{\small{}1} & \textcolor{green}{\small{}-167}&\textcolor{green}{\small{}8} & \textcolor{green}{\small{}-108}&\textcolor{green}{\small{}7}\tabularnewline
\textbf{\small{}$M_{4}$} & \textcolor{purple}{\small{}33}&\textcolor{purple}{\small{}9} & \textcolor{purple}{\small{}39}&\textcolor{purple}{\small{}3} & \textcolor{purple}{\small{}-8}&\textcolor{purple}{\small{}9} & \multicolumn{2}{c}{{\small{}---}} & \textcolor{green}{\small{}33}&\textcolor{green}{\small{}1} & \textcolor{green}{\small{}-351}&\textcolor{green}{\small{}1} & \textcolor{green}{\small{}219}&\textcolor{green}{\small{}3} & \textcolor{green}{\small{}512}&\textcolor{green}{\small{}8} & \textcolor{green}{\small{}444}&\textcolor{green}{\small{}6} & \textcolor{green}{\small{}446}&\textcolor{green}{\small{}5} & \textcolor{green}{\small{}-424}&\textcolor{green}{\small{}1} & \textcolor{green}{\small{}-316}&\textcolor{green}{\small{}1} & \textcolor{green}{\small{}24}&\textcolor{green}{\small{}5} & \textcolor{green}{\small{}-111}&\textcolor{green}{\small{}2} & \textcolor{green}{\small{}-52}&\textcolor{green}{\small{}1}\tabularnewline
\textbf{\small{}$M_{5}$} & \textcolor{purple}{\small{}48}&\textcolor{purple}{\small{}5} & \textcolor{purple}{\small{}53}&\textcolor{purple}{\small{}9} & \textcolor{purple}{\small{}5}&\textcolor{purple}{\small{}7} & \textcolor{purple}{\small{}14}&\textcolor{purple}{\small{}6} & \multicolumn{2}{c}{{\small{}---}} & \textcolor{green}{\small{}-384}&\textcolor{green}{\small{}2} & \textcolor{green}{\small{}186}&\textcolor{green}{\small{}2} & \textcolor{green}{\small{}479}&\textcolor{green}{\small{}7} & \textcolor{green}{\small{}411}&\textcolor{green}{\small{}5} & \textcolor{green}{\small{}413}&\textcolor{green}{\small{}5} & \textcolor{green}{\small{}-457}&\textcolor{green}{\small{}2} & \textcolor{green}{\small{}-349}&\textcolor{green}{\small{}2} & \textcolor{green}{\small{}-8}&\textcolor{green}{\small{}6} & \textcolor{green}{\small{}-144}&\textcolor{green}{\small{}3} & \textcolor{green}{\small{}-85}&\textcolor{green}{\small{}2}\tabularnewline
\textbf{\small{}$M_{6}$} & \textcolor{purple}{\small{}0}&\textcolor{purple}{\small{}0} & \textcolor{purple}{\small{}5}&\textcolor{purple}{\small{}4} & \textcolor{purple}{\small{}-42}&\textcolor{purple}{\small{}8} & \textcolor{purple}{\small{}-33}&\textcolor{purple}{\small{}9} & \textcolor{purple}{\small{}-48}&\textcolor{purple}{\small{}5} & \multicolumn{2}{c}{{\small{}---}} & \textcolor{green}{\small{}570}&\textcolor{green}{\small{}4} & \textcolor{green}{\small{}863}&\textcolor{green}{\small{}9} & \textcolor{green}{\small{}795}&\textcolor{green}{\small{}7} & \textcolor{green}{\small{}797}&\textcolor{green}{\small{}7} & \textcolor{green}{\small{}-73}&\textcolor{green}{\small{}0} & \textcolor{green}{\small{}35}&\textcolor{green}{\small{}0} & \textcolor{green}{\small{}375}&\textcolor{green}{\small{}6} & \textcolor{green}{\small{}240}&\textcolor{green}{\small{}0} & \textcolor{green}{\small{}299}&\textcolor{green}{\small{}0}\tabularnewline
\textbf{\small{}$M_{7}$} & \textcolor{purple}{\small{}57}&\textcolor{purple}{\small{}8} & \textcolor{purple}{\small{}63}&\textcolor{purple}{\small{}3} & \textcolor{purple}{\small{}15}&\textcolor{purple}{\small{}1} & \textcolor{purple}{\small{}24}&\textcolor{purple}{\small{}0} & \textcolor{purple}{\small{}9}&\textcolor{purple}{\small{}4} & \textcolor{purple}{\small{}57}&\textcolor{purple}{\small{}9} & \multicolumn{2}{c}{{\small{}---}} & \textcolor{green}{\small{}293}&\textcolor{green}{\small{}5} & \textcolor{green}{\small{}225}&\textcolor{green}{\small{}3} & \textcolor{green}{\small{}227}&\textcolor{green}{\small{}2} & \textcolor{green}{\small{}-643}&\textcolor{green}{\small{}4} & \textcolor{green}{\small{}-535}&\textcolor{green}{\small{}4} & \textcolor{green}{\small{}-194}&\textcolor{green}{\small{}8} & \textcolor{green}{\small{}-330}&\textcolor{green}{\small{}5} & \textcolor{green}{\small{}-271}&\textcolor{green}{\small{}4}\tabularnewline
\textbf{\small{}$M_{8}$} & \textcolor{purple}{\small{}49}&\textcolor{purple}{\small{}2} & \textcolor{purple}{\small{}54}&\textcolor{purple}{\small{}7} & \textcolor{purple}{\small{}6}&\textcolor{purple}{\small{}4} & \textcolor{purple}{\small{}15}&\textcolor{purple}{\small{}4} & \textcolor{purple}{\small{}0}&\textcolor{purple}{\small{}8} & \textcolor{purple}{\small{}49}&\textcolor{purple}{\small{}3} & \textcolor{purple}{\small{}-8}&\textcolor{purple}{\small{}6} & \multicolumn{2}{c}{{\small{}---}} & \textcolor{green}{\small{}-68}&\textcolor{green}{\small{}3} & \textcolor{green}{\small{}-66}&\textcolor{green}{\small{}3} & \textcolor{green}{\small{}-936}&\textcolor{green}{\small{}9} & \textcolor{green}{\small{}-828}&\textcolor{green}{\small{}9} & \textcolor{green}{\small{}-488}&\textcolor{green}{\small{}3} & \multicolumn{2}{c}{\textcolor{green}{\small{}-624}} & \textcolor{green}{\small{}-564}&\textcolor{green}{\small{}9}\tabularnewline
\textbf{\small{}$M_{9}$} & \textcolor{purple}{\small{}72}&\textcolor{purple}{\small{}4} & \textcolor{purple}{\small{}77}&\textcolor{purple}{\small{}9} & \textcolor{purple}{\small{}29}&\textcolor{purple}{\small{}7} & \textcolor{purple}{\small{}38}&\textcolor{purple}{\small{}6} & \textcolor{purple}{\small{}24}&\textcolor{purple}{\small{}0} & \textcolor{purple}{\small{}72}&\textcolor{purple}{\small{}5} & \textcolor{purple}{\small{}14}&\textcolor{purple}{\small{}6} & \textcolor{purple}{\small{}23}&\textcolor{purple}{\small{}2} & \multicolumn{2}{c}{{\small{}---}} & \textcolor{green}{\small{}2}&\textcolor{green}{\small{}0} & \textcolor{green}{\small{}-868}&\textcolor{green}{\small{}6} & \textcolor{green}{\small{}-760}&\textcolor{green}{\small{}6} & \textcolor{green}{\small{}-420}&\textcolor{green}{\small{}1} & \textcolor{green}{\small{}-555}&\textcolor{green}{\small{}7} & \textcolor{green}{\small{}-496}&\textcolor{green}{\small{}7}\tabularnewline
\textbf{\small{}$M_{10}$} & \textcolor{purple}{\small{}69}&\textcolor{purple}{\small{}4} & \textcolor{purple}{\small{}74}&\textcolor{purple}{\small{}8} & \textcolor{purple}{\small{}26}&\textcolor{purple}{\small{}6} & \textcolor{purple}{\small{}35}&\textcolor{purple}{\small{}5} & \textcolor{purple}{\small{}20}&\textcolor{purple}{\small{}9} & \textcolor{purple}{\small{}69}&\textcolor{purple}{\small{}4} & \textcolor{purple}{\small{}11}&\textcolor{purple}{\small{}5} & \textcolor{purple}{\small{}20}&\textcolor{purple}{\small{}1} & \textcolor{purple}{\small{}-3}&\textcolor{purple}{\small{}1} & \multicolumn{2}{c}{{\small{}---}} & \textcolor{green}{\small{}-870}&\textcolor{green}{\small{}6} & \textcolor{green}{\small{}-762}&\textcolor{green}{\small{}6} & \multicolumn{2}{c}{\textcolor{green}{\small{}-422}} & \textcolor{green}{\small{}-557}&\textcolor{green}{\small{}7} & \textcolor{green}{\small{}-498}&\textcolor{green}{\small{}6}\tabularnewline
\textbf{\small{}$M_{11}$} & \textcolor{purple}{\small{}-637}&\textcolor{purple}{\small{}1} & \textcolor{purple}{\small{}-631}&\textcolor{purple}{\small{}6} & \textcolor{purple}{\small{}-679}&\textcolor{purple}{\small{}8} & \textcolor{purple}{\small{}-670}&\textcolor{purple}{\small{}9} & \textcolor{purple}{\small{}-685}&\textcolor{purple}{\small{}5} & \multicolumn{2}{c}{\textcolor{purple}{\small{}-637}} & \textcolor{purple}{\small{}-694}&\textcolor{purple}{\small{}9} & \textcolor{purple}{\small{}-686}&\textcolor{purple}{\small{}3} & \textcolor{purple}{\small{}-709}&\textcolor{purple}{\small{}5} & \textcolor{purple}{\small{}-706}&\textcolor{purple}{\small{}4} & \multicolumn{2}{c}{{\small{}---}} & \textcolor{green}{\small{}108}&\textcolor{green}{\small{}0} & \textcolor{green}{\small{}448}&\textcolor{green}{\small{}6} & \textcolor{green}{\small{}312}&\textcolor{green}{\small{}9} & \textcolor{green}{\small{}372}&\textcolor{green}{\small{}0}\tabularnewline
\textbf{\small{}$M_{12}$} & \textcolor{purple}{\small{}95}&\textcolor{purple}{\small{}6} & \textcolor{purple}{\small{}101}&\textcolor{purple}{\small{}1} & \textcolor{purple}{\small{}52}&\textcolor{purple}{\small{}9} & \textcolor{purple}{\small{}61}&\textcolor{purple}{\small{}8} & \textcolor{purple}{\small{}47}&\textcolor{purple}{\small{}2} & \textcolor{purple}{\small{}95}&\textcolor{purple}{\small{}7} & \textcolor{purple}{\small{}37}&\textcolor{purple}{\small{}8} & \textcolor{purple}{\small{}46}&\textcolor{purple}{\small{}4} & \textcolor{purple}{\small{}23}&\textcolor{purple}{\small{}2} & \textcolor{purple}{\small{}26}&\textcolor{purple}{\small{}3} & \textcolor{purple}{\small{}732}&\textcolor{purple}{\small{}7} & \multicolumn{2}{c}{{\small{}---}} & \textcolor{green}{\small{}340}&\textcolor{green}{\small{}6} & \textcolor{green}{\small{}204}&\textcolor{green}{\small{}9} & \textcolor{green}{\small{}264}&\textcolor{green}{\small{}0}\tabularnewline
\textbf{\small{}$M_{13}$} & \textcolor{purple}{\small{}63}&\textcolor{purple}{\small{}9} & \textcolor{purple}{\small{}69}&\textcolor{purple}{\small{}3} & \textcolor{purple}{\small{}21}&\textcolor{purple}{\small{}1} & \multicolumn{2}{c}{\textcolor{purple}{\small{}30}} & \textcolor{purple}{\small{}15}&\textcolor{purple}{\small{}4} & \textcolor{purple}{\small{}63}&\textcolor{purple}{\small{}9} & \textcolor{purple}{\small{}6}&\textcolor{purple}{\small{}0} & \textcolor{purple}{\small{}14}&\textcolor{purple}{\small{}6} & \textcolor{purple}{\small{}-8}&\textcolor{purple}{\small{}6} & \textcolor{purple}{\small{}-5}&\textcolor{purple}{\small{}5} & \textcolor{purple}{\small{}700}&\textcolor{purple}{\small{}9} & \textcolor{purple}{\small{}-31}&\textcolor{purple}{\small{}8} & \multicolumn{2}{c}{{\small{}---}} & \textcolor{green}{\small{}-135}&\textcolor{green}{\small{}7} & \textcolor{green}{\small{}-76}&\textcolor{green}{\small{}6}\tabularnewline
\textbf{\small{}$M_{14}$} & \textcolor{purple}{\small{}78}&\textcolor{purple}{\small{}7} & \textcolor{purple}{\small{}84}&\textcolor{purple}{\small{}1} & \textcolor{purple}{\small{}35}&\textcolor{purple}{\small{}9} & \textcolor{purple}{\small{}44}&\textcolor{purple}{\small{}9} & \textcolor{purple}{\small{}30}&\textcolor{purple}{\small{}2} & \textcolor{purple}{\small{}78}&\textcolor{purple}{\small{}7} & \textcolor{purple}{\small{}20}&\textcolor{purple}{\small{}9} & \textcolor{purple}{\small{}29}&\textcolor{purple}{\small{}5} & \textcolor{purple}{\small{}6}&\textcolor{purple}{\small{}3} & \textcolor{purple}{\small{}9}&\textcolor{purple}{\small{}3} & \textcolor{purple}{\small{}715}&\textcolor{purple}{\small{}8} & \textcolor{purple}{\small{}-16}&\textcolor{purple}{\small{}9} & \textcolor{purple}{\small{}14}&\textcolor{purple}{\small{}8} & \multicolumn{2}{c}{{\small{}---}} & \textcolor{green}{\small{}59}&\textcolor{green}{\small{}1}\tabularnewline
\textbf{\small{}$M_{15}$} & \textcolor{purple}{\small{}80}&\textcolor{purple}{\small{}2} & \textcolor{purple}{\small{}85}&\textcolor{purple}{\small{}7} & \textcolor{purple}{\small{}37}&\textcolor{purple}{\small{}4} & \textcolor{purple}{\small{}46}&\textcolor{purple}{\small{}4} & \textcolor{purple}{\small{}31}&\textcolor{purple}{\small{}8} & \textcolor{purple}{\small{}80}&\textcolor{purple}{\small{}3} & \textcolor{purple}{\small{}22}&\textcolor{purple}{\small{}4} & \textcolor{purple}{\small{}31}&\textcolor{purple}{\small{}0} & \textcolor{purple}{\small{}7}&\textcolor{purple}{\small{}8} & \textcolor{purple}{\small{}10}&\textcolor{purple}{\small{}9} & \textcolor{purple}{\small{}717}&\textcolor{purple}{\small{}3} & \textcolor{purple}{\small{}-15}&\textcolor{purple}{\small{}4} & \textcolor{purple}{\small{}16}&\textcolor{purple}{\small{}4} & \textcolor{purple}{\small{}1}&\textcolor{purple}{\small{}5} & \multicolumn{2}{c}{{\small{}---}}\tabularnewline
\hline 
\end{tabular}{\small\par}
\end{turn}
\par\end{centering}
\caption{Detailed model selection statistics $\widehat{\mathrm{elpd}}_{CV}\left(M_{A},M_{B}\,|\,y\right)$
for all models considered in Example~1. Positive numbers favor models
labeled on rows (lhs), negative numbers favor columns (top). \textcolor{green}{Green
statistics} are computed jointly (i.e. $\widehat{\mathrm{elpd}}_{CV}^{j}\left(M_{A},M_{B}\,|\,y\right)$),
\textcolor{purple}{purple statistics} computed pointwise (i.e. $\widehat{\mathrm{elpd}}_{CV}^{pw}\left(M_{A},M_{B}\,|\,y\right)$).}
\end{table}

\subsection{Lung cancer in Pennsylvania}

\begin{figure}[H]
\begin{centering}
\includegraphics[width=1\textwidth]{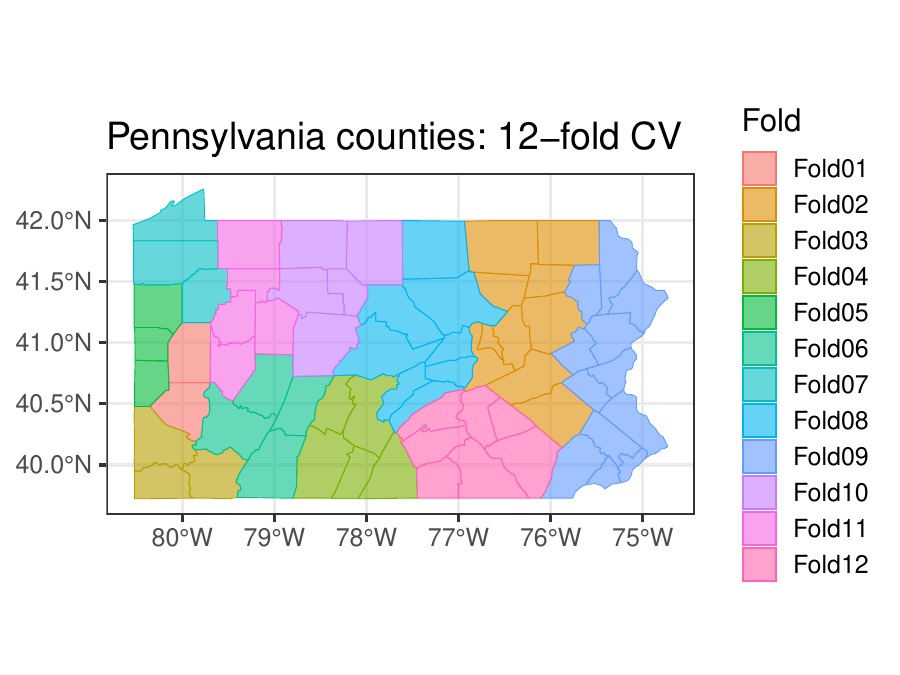}
\par\end{centering}
\caption{Spatially-clustered CV folds computed using the spatialsample package
\parencite{mahoneyAssessingPerformanceSpatial2023}.}
\label{fig:penn-clust}
\end{figure}

The three model forms used are:\label{list:modelforms}
\begin{description}
\item [{BYM}] Besag-York-Molli\'e model \parencite{besagBayesianImageRestoration1991}
\item [{BYM2}] Re-parameterized BYM model \parencite{rieblerIntuitiveBayesianSpatial2016}
\item [{SAR}] Simultaneous autoregression \cite{anselin1988spatial}
\end{description}

\begin{table}[H]
\begin{centering}
\begin{tabular}{lccccccccc}
\hline 
 &  & \multicolumn{8}{c}{\textbf{Candidate model}}\tabularnewline
 &  & $M_1$ & $M_2$ & $M_3$ & $M_4$ & $M_5$ & $M_6$ & $M_7$ & $M_8$\tabularnewline
\cline{1-1} \cline{3-10} \cline{4-10} \cline{5-10} \cline{6-10} \cline{7-10} \cline{8-10} \cline{9-10} \cline{10-10} 
Model form &  & BYM2 & BYM2 & BYM & BYM & SAR & SAR & SAR & SAR\tabularnewline
Weights &  & $W$ & $W$ & $W$ & $W$ & $W$ & $W$ & $W_{+}$ & $W_{+}$\tabularnewline
Covariate &  &  &  &  &  &  &  &  & \tabularnewline
\quad{}Smoking &  & \Checkmark{} &  & \Checkmark{} &  & \Checkmark{} &  & \Checkmark{} & \tabularnewline
\hline 
\end{tabular}
\par\end{centering}
\caption{Candidate models for the Pennsylvania lung cancer example. The model form
abbreviations refer to the list on page~\pageref{list:modelforms}.}

\label{tbl:candidates}
\end{table}

\begin{table}[H]
\begin{centering}
\begin{tabular}{llr@{\extracolsep{0pt}.}lr@{\extracolsep{0pt}.}lr@{\extracolsep{0pt}.}lr@{\extracolsep{0pt}.}lr@{\extracolsep{0pt}.}l}
\hline 
Parameter &  & \multicolumn{2}{c}{mean} & \multicolumn{2}{c}{sd} & \multicolumn{2}{c}{$q_{0.025}$} & \multicolumn{2}{c}{$q_{0.5}$} & \multicolumn{2}{c}{$q_{0.975}$}\tabularnewline
\cline{1-1} \cline{3-12} \cline{5-12} \cline{7-12} \cline{9-12} \cline{11-12} 
\noalign{\vskip6pt}
Precision ($\tau$) &  & 127&8 & 51&3 & 56&9 & 118&1 & 255&2\tabularnewline
Spatial parameter ($\rho$) &  & 0&58 & 0&068 & 0&44 & 0&58 & 0&71\tabularnewline
Regression coef. ($\beta$) &  & \multicolumn{2}{c}{} & \multicolumn{2}{c}{} & \multicolumn{2}{c}{} & \multicolumn{2}{c}{} & \multicolumn{2}{c}{}\tabularnewline
\quad{}(Intercept) &  & -0&054 & 0&022 & -0&099 & -0&053 & -0&013\tabularnewline
\hline 
\end{tabular}
\par\end{centering}
\caption{Summary of posterior marginals for the preferred model for the Pennsylvania
lung cancer example (Section~4.2, main text). The preferred
model does not include smoking as a regression coefficient. Note the spatial
parameter $\rho=0.58$ does not suggest differences between joint and pointwise
CV will be large.}

\label{penn-prefer-coef}
\end{table}

\begin{table}[H]
\begin{centering}
\begin{tabular}{ccr@{\extracolsep{0pt}.}lr@{\extracolsep{0pt}.}lr@{\extracolsep{0pt}.}lr@{\extracolsep{0pt}.}lr@{\extracolsep{0pt}.}lr@{\extracolsep{0pt}.}lr@{\extracolsep{0pt}.}lr@{\extracolsep{0pt}.}l}
\hline 
 &  & \multicolumn{2}{c}{$M_{1}$} & \multicolumn{2}{c}{$M_{2}$} & \multicolumn{2}{c}{$M_{3}$} & \multicolumn{2}{c}{$M_{4}$} & \multicolumn{2}{c}{$M_{5}$} & \multicolumn{2}{c}{$M_{6}$} & \multicolumn{2}{c}{$M_{7}$} & \multicolumn{2}{c}{$M_{8}$}\tabularnewline
\hline 
$M_{1}$ &  & \multicolumn{2}{c}{---} & \textcolor{green}{-0}&\textcolor{green}{4} & \textcolor{green}{4}&\textcolor{green}{6} & \textcolor{green}{5}&\textcolor{green}{0} & \textcolor{green}{-0}&\textcolor{green}{4} & \textcolor{green}{-0}&\textcolor{green}{6} & \textcolor{green}{-0}&\textcolor{green}{3} & \textcolor{green}{-0}&\textcolor{green}{7}\tabularnewline
$M_{2}$ &  & \textcolor{purple}{0}&\textcolor{purple}{4} & \multicolumn{2}{c}{---} & \textcolor{green}{4}&\textcolor{green}{9} & \textcolor{green}{5}&\textcolor{green}{4} & \textcolor{green}{0}&\textcolor{green}{0} & \textcolor{green}{-0}&\textcolor{green}{2} & \textcolor{green}{0}&\textcolor{green}{0} & \textcolor{green}{-0}&\textcolor{green}{3}\tabularnewline
$M_{3}$ &  & \textcolor{purple}{-4}&\textcolor{purple}{5} & \textcolor{purple}{-4}&\textcolor{purple}{8} & \multicolumn{2}{c}{---} & \textcolor{green}{0}&\textcolor{green}{5} & \textcolor{green}{-5}&\textcolor{green}{0} & \textcolor{green}{-5}&\textcolor{green}{2} & \textcolor{green}{-4}&\textcolor{green}{9} & \textcolor{green}{-5}&\textcolor{green}{3}\tabularnewline
$M_{4}$ &  & \textcolor{purple}{-4}&\textcolor{purple}{9} & \textcolor{purple}{-5}&\textcolor{purple}{3} & \textcolor{purple}{-0}&\textcolor{purple}{5} & \multicolumn{2}{c}{---} & \textcolor{green}{-5}&\textcolor{green}{4} & \textcolor{green}{-5}&\textcolor{green}{6} & \textcolor{green}{-5}&\textcolor{green}{4} & \textcolor{green}{-5}&\textcolor{green}{7}\tabularnewline
$M_{5}$ &  & \textcolor{purple}{0}&\textcolor{purple}{3} & \textcolor{purple}{-0}&\textcolor{purple}{1} & \textcolor{purple}{4}&\textcolor{purple}{7} & \textcolor{purple}{5}&\textcolor{purple}{2} & \multicolumn{2}{c}{---} & \textcolor{green}{-0}&\textcolor{green}{2} & \textcolor{green}{0}&\textcolor{green}{1} & \textcolor{green}{-0}&\textcolor{green}{3}\tabularnewline
$M_{6}$ &  & \textcolor{purple}{0}&\textcolor{purple}{5} & \textcolor{purple}{0}&\textcolor{purple}{1} & \textcolor{purple}{5}&\textcolor{purple}{0} & \textcolor{purple}{5}&\textcolor{purple}{4} & \textcolor{purple}{0}&\textcolor{purple}{2} & \multicolumn{2}{c}{---} & \textcolor{green}{0}&\textcolor{green}{2} & \textcolor{green}{-0}&\textcolor{green}{1}\tabularnewline
$M_{7}$ &  & \textcolor{purple}{0}&\textcolor{purple}{3} & \textcolor{purple}{-0}&\textcolor{purple}{1} & \textcolor{purple}{4}&\textcolor{purple}{7} & \textcolor{purple}{5}&\textcolor{purple}{2} & \textcolor{purple}{0}&\textcolor{purple}{0} & \textcolor{purple}{-0}&\textcolor{purple}{2} & \multicolumn{2}{c}{---} & \textcolor{green}{-0}&\textcolor{green}{4}\tabularnewline
$M_{8}$ &  & \textcolor{purple}{0}&\textcolor{purple}{6} & \textcolor{purple}{0}&\textcolor{purple}{2} & \textcolor{purple}{5}&\textcolor{purple}{0} & \textcolor{purple}{5}&\textcolor{purple}{5} & \textcolor{purple}{0}&\textcolor{purple}{3} & \textcolor{purple}{0}&\textcolor{purple}{1} & \textcolor{purple}{0}&\textcolor{purple}{3} & \multicolumn{2}{c}{---}\tabularnewline
\hline 
\end{tabular}
\par\end{centering}
\caption{Selection statistics $\widehat{\mathrm{elpd}}_{CV}\left(M_{A},M_{B}\,|\,y\right)$
for all models in Example~2. Positive numbers favor models
labeled on rows (lhs), negative numbers favor columns (top). \textcolor{green}{Green
statistics} are computed jointly,
\textcolor{purple}{purple statistics} computed pointwise.}
\end{table}

\section{Laplace approximation}\label{sec:laplace}

The simulation studies in Section~3 conduct fast approximate Bayesian inference by 
Laplace approximation \parencite{mackay2003information}. Fast, approximate inference
is required because the simulation study fits millions of partial-data posteriors.
Here we briefly describe the approach and illustrate the accuracy of the resulting posteriors.

To implement CV, we need to conduct inference in a way that leaves certain data elements out of the
training set. To do this, we regard the left-out data (i.e. the test set and buffer) as random variables,
rather than the fixed observations that appear in the original $y$. We will replace
$y_{\mathsf{test}}$ and $y_{\mathsf{buffer}}$ with $y_{\mathsf{test}}^*$ and $y_{\mathsf{buffer}}^*$, 
respectively, where the superscript $y^*$ denotes a vector of the same shape as the original.

Since the models in Section~3 of the text have linear link functions and Gaussian observation
densities, the states $z$ can be marginalized out analytically. The resulting joint log density
is of the form
\begin{equation}
    \log p\left(\theta,y_{\mathsf{test}_{k}}^*,y_{\mathsf{buffer}_{k}}^*,y_{\mathsf{train}_{k}}\right)=\log p\left(\theta\right)+\log\mathcal{N}\left(\begin{pmatrix}y_{\mathsf{train}_{k}}\\
y_{\mathsf{buffer}_{k}}^*\\
y_{\mathsf{test}_{k}}^*
\end{pmatrix}\,|\,\mu_{\theta},\Lambda_{\theta}^{-1}\right)
\label{eq:joint}
\end{equation}
where the notation assumes a suitable ordering of the data, mean, and precision
so that the buffer and test vectors are appended to the test set. 

The structure of \eqref{eq:joint} allows JAX \parencite{jax2018github} to vectorize
computations across multiple posterior fits, so long as the size of the training, buffer,
and test vectors have the same shape.

Laplace approximation proceeds by finding the maximum a postiori (MAP) estimate by
optimizing,
\begin{equation}
\left(\widehat\theta,\widehat{y_{\mathsf{test}_{k}}},\widehat{y_{\mathsf{buffer}_{k}}}\right)
:= \argmax_{\theta,y_{\mathsf{test}_{k}^*},y_{\mathsf{buffer}_{k}^*}} \log p\left(\theta,y_{\mathsf{test}_{k}}^*,y_{\mathsf{buffer}_{k}}^*,y_{\mathsf{train}_{k}}\right),
\end{equation}
which we find by L-BFGS implemented in \texttt{jaxopt} \parencite{jaxopt_implicit_diff}.

To make the optimization reliable, we transform the parameters in \eqref{eq:joint} using
bijectors provided by Tensorflow Probability \parencite{dillon2017tensorflow}, and apply the associated log Jacobian determinant adjustments. Bounded parameters
are transformed using a sigmoidal transformation and positive parameters use softplus.

The inverse covariance matrix for the parameter and test set is found by computing the negative Hessian 
(second derivative) matrix of \eqref{eq:joint} evaluated at the MAP,
a computation that takes advantage of JAX's automatic differentiation
features.

\begin{figure}[H]
\includegraphics[width=1\textwidth]{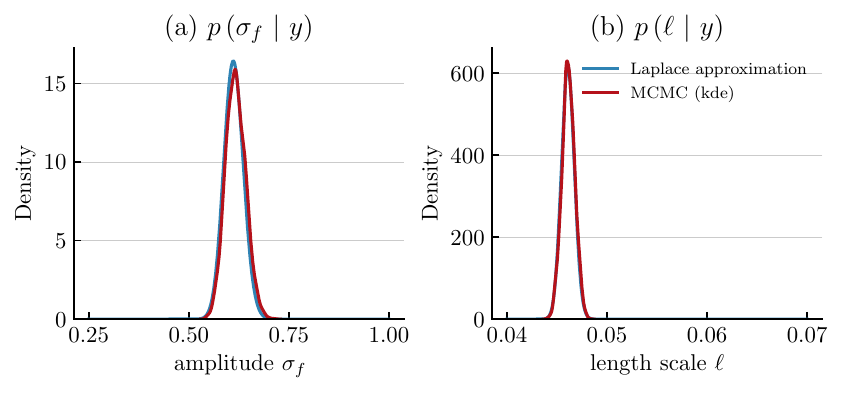}
\caption{Comparison of marginal posterior distributions computed by Laplace approximation (blue)
    and MCMC (red) for a single data draw of the exponentiated quadratic kernel model
    (Section~3.2, main text).
    The MCMC marginal is a kernel density estimate estimated with a Gaussian kernel. MCMC is performed using \texttt{blackjax}'s \parencite{cabezas2024blackjax} No-U-Turn
    Sampler \parencite{hoffman2014no} implementation with 4,000 draws across 4 independent chains.}
\label{fig:marginal:eq}
\end{figure}

\begin{figure}[H]
\includegraphics[width=1\textwidth]{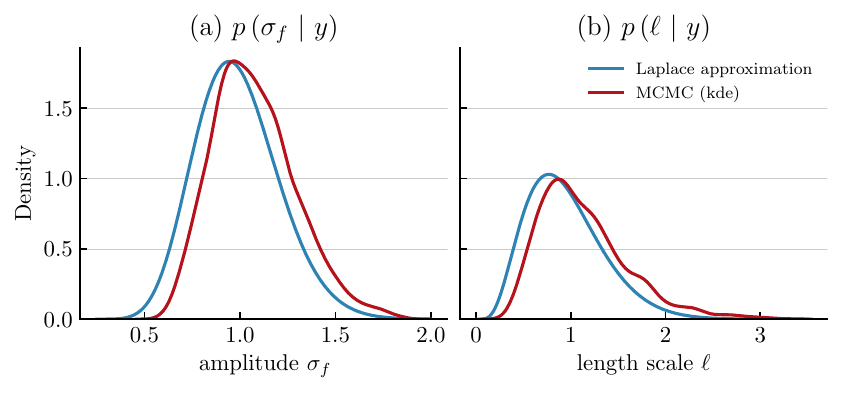}
\caption{Comparison of marginal posterior distributions computed by Laplace approximation (blue)
    and MCMC (red) for a single data draw of the Mat\'ern kernel model (Section~3.2, main text).
    The MCMC marginal is a 
    kernel density estimate estimated with a Gaussian kernel. MCMC is performed using \texttt{blackjax}'s \parencite{cabezas2024blackjax} No-U-Turn
    Sampler \parencite{hoffman2014no} implementation with 4,000 draws across 4 independent chains.}
\label{fig:marginal:mat}
\end{figure}

The resulting posterior distributions appear similar to equivalent accurate inference
conducted using MCMC. Figures~\ref{fig:marginal:eq} and \ref{fig:marginal:mat} below provide examples
for a single data draw for the simulation in Section~3.2 in the text.

%\printbibliography[heading=bibintoc]

\end{document}